\documentclass[aps,preprint]{revtex4}
\usepackage{amsmath}
\usepackage{amssymb}
\usepackage{graphicx}
\usepackage{bm}
\usepackage{mathrsfs}
\usepackage{color}
\usepackage{subfigure}
\usepackage{epstopdf, epsfig}
\usepackage{array} 
\usepackage{stackengine}
\usepackage[normalem]{ulem}

\usepackage{dcolumn}% Align table columns on decimal point (pls)
\usepackage{soul}% barrer (pls)
\usepackage[dvipsnames]{xcolor}
\stackMath

\usepackage{natbib}

\usepackage{booktabs}
\usepackage{multirow}

\newcommand{\beq}{\begin{equation}}
\newcommand{\eeq}{\end{equation}}

\newcommand{\bpar}{B_\parallel}
\newcommand{\apar}{ A_{\parallel}}

\newcommand{\lapp}{\nabla_{\perp}^2}

\newcommand{\pa}{\partial}

\newcommand{\nno}{\nonumber}

\newcommand{\ben}{\begin{eqnarray}}
\newcommand{\een}{\end{eqnarray}}

\newcommand{\bxi}{\bar{\xi}}

\newcommand{\lappinv}{\Delta_{\perp}^{-1}}

\newcommand{\tphi}{\tilde{\phi}}
\newcommand{\tapar}{\tilde{A}}
\newcommand{\aeq}{\apar^{(0)}}
\newcommand{\txi}{\tilde{\xi}_{in}}
\newcommand{\ba}{\bar{a}}

\begin{document}

\title{Influence of ion-to-electron temperature ratio on tearing instability and resulting subion-scale turbulence in a low-$\beta_e$ collisionless plasma}

\author{C.Granier$^{1,*}$, E. Tassi$^{2}$, D. Laveder$^{2}$, T. Passot$^{2}$, P.L. Sulem$^{2}$}
\affiliation{$^{1)}$ Max Planck Institute for Plasma Physics, Boltzmannstraße 2, 85748 Garching,
Germany\\
$^{2)}$Universit\'e C\^ote d'Azur, CNRS, Observatoire de la C\^ote d'Azur, Laboratoire J.L. Lagrange, Boulevard de l'Observatoire, CS  34229, 06304 Nice Cedex 4, France\\
$*$ Corresponding author: camille.granier@ipp.mpg.de}
\date{\today}

\baselineskip 24pt

\begin{abstract}
A two-field gyrofluid model including ion finite Larmor radius (FLR) corrections, magnetic fluctuations along the ambient field and electron inertia is used to study two-dimensional reconnection in a low $\beta_e$ collisionless plasma, in a plane perpendicular to the ambient field.  Both moderate and large values of the ion-to-electron temperature ratio $\tau$ are considered. The linear growth rate of the tearing instability is computed for various values of $\tau$, confirming the convergence to reduced electron magnetohydrodynamics (REMHD) predictions in the large $\tau$ limit. Comparisons with analytical estimates in several limit cases are also presented. The nonlinear dynamics leads to a fully-developed turbulent regime that appears to be sensitive to the value of the parameter $\tau$. For $\tau = 100$, strong large-scale velocity shears trigger Kelvin-Helmholtz instability, leading to the propagation of the turbulence  through the separatrices, together with the formation of eddies of size of the order of the electron skin depth. In the  $\tau = 1$ regime, the vortices are significantly smaller and their accurate description requires that electron FLR effects be taken into account.
\end{abstract}

\maketitle

\section{Introduction}

Magnetic reconnection plays an important role in various space-plasma phenomena, from solar flares to geomagnetic storms. The investigation of magnetic reconnection has provided some crucial understanding of the mechanisms responsible for the release of energy particularly in the context of astrophysical plasmas, where the collisional mean free path is large enough for classical Coulomb collisions to be negligible. It is now acknowledged that reconnection in nature is often driven by collisionless effects. This necessitates models capable of including two-fluid effects, such as electron inertia.  A significant step forward was taken in Refs. \cite{Ayd92, Ott93}, where it was shown that, in the collisionless regime, two-fluid effects driving reconnection can provide a way to achieve fast reconnection. Subsequent bodies of work have confirmed the crucial role played by collisionless effects (see for example Refs. \cite{Bis97, Gra00,  Wan00, Bir01, Gra10, Fit07, Num15}), and have shown a good agreement with in situ spacecraft measurements in the Earth’s magnetosphere \cite{Che17}. Notably, some of these studies involved fully kinetic simulations (as, for instance, in Ref. \cite{Ege19}). However, global fully kinetic simulations usually remain extremely expensive, and simplified models or hybrid approaches have emerged as alternatives with the potential to efficiently capture essential physical phenomena while significantly reducing computational costs. In the presence of a strong ambient magnetic field component, known as the 'guide field', gyrokinetic and gyrofluid models hold great potential for reconnection simulations, as shown in Refs. \cite{Rog07, Com13, Zac14, Num15, Tas18}.

In this paper, we make use of a two-field gyrofluid model derived in Ref. \cite{Pas18} to simulate numerically reconnection events driven by electron inertia. This model isolates the dynamics of Alfv\'en waves (at the magnetohydrodynamics (MHD) scales) and kinetic Alfv\'en waves (at the sub-ion scales) in regimes where the couplings to the other kinds of waves are subdominant. It provides a good toolset for analyzing the plasma behavior in the strong guide field, low-$\beta_e$ regime, where $\beta_e$ is the ratio between electron kinetic pressure and guide field magnetic pressure. 
The model includes ion Larmor radius effects, and enables an arbitrary equilibrium ion-to-electron temperature ratio $\tau$. For instance, this model is particularly valuable for investigating reconnection phenomena in regions such as the solar corona and its vicinity ($\beta_i = \tau \beta_e \lesssim 0.1$), the Earth's magnetosheath (where $\beta_i \lesssim 1$ can be observed), and the Earth's magnetosphere ($\beta_i \ll 0.1$) \cite{Tre13}. Note that observations have revealed that, in most astrophysical plasmas, ion temperatures are usually larger than those of electrons, with for example $\tau \gtrsim 10$ in the Earth's magnetosphere (see  \cite{Bur16} for a measurement of magnetic reconnection at the magnetopause),  $\tau \sim 3 - 4$ in the Earth's  magnetosheath \cite{Eas18} and $\tau \sim 2$ in the solar wind \cite{Per17}.
It thus appears relevant to investigate the effect of the temperature ratio within the framework of  the two-field gyrofluid model, which is computationally less demanding than the kinetic descriptions.
This model  bridges the gap between reduced magnetohydrodynamics (RMHD), the inertial kinetic Alfv\'en waves (IKAW) model \cite{Che17, Pas17, Pas19}, and a reduced electron magnetohydrodynamics (REMHD) model that accounts for electron inertia, distinguishing it from the REMHD model derived in Ref. \cite{Sch09}. In two spatial dimensions, the REMHD equations are formally identical to those of the electron magnetohydrodynamics (EMHD) model \cite{Kin90,Bis00} which focuses on the incompressible regime, describing whistler waves.

The present work concentrates on the two-dimensional dynamics that develops in a plane perpendicular to the ambient field. Its aim  is twofold. We first examine the linear growth rates of the tearing instability, investigating various equilibrium temperature ratios  and confirming that the model converges toward the REMHD regime as $\tau$ increases. We point out that, compared to previous investigations on the role of the ion thermal radius and based on gyrofluid models \cite{Gra00,Com12,Com13}, the present analysis does not require $\beta_i \ll1$, where $\beta_i= \tau \beta_e$. Relaxing this assumption makes it possible to access the above mentioned IKAW and REMHD regimes. We then study the turbulence regime resulting from reconnection in the cases of moderate and large values of the $\tau$ parameter.

Previous numerical simulations  conducted in the cold-ion regime, using a reduced description that appears as a limit of our model, have provided evidence that collisionless magnetic reconnection can trigger fluid-like secondary instabilities \cite{Del03, Del05,Del06, Del11, Gra07}. Always in the cold-ion limit, fluid-like secondary instabilities were observed in two and three-dimensional numerical simulations of a four-field model accounting for finite $\beta_e$ effects \cite{Gra09,Tas10,Gra12} It was observed that for low-$\beta_e$ reconnection, Kelvin-Helmholtz or Rayleigh-Taylor-like instabilities can develop, depending on the ratio of the ion-sound Larmor radius and electron skin depth $\rho_s/d_e$. These secondary instabilities could potentially act as a source of turbulence. In the present study, we consider a broad range of values for the parameter $\tau$ and analyze the  influence of this parameter on  the nonlinear evolution of magnetic islands,  as well as on the properties of the turbulence driven by  the secondary instabilities. Nonlinear simulations were done for two distinct ion-to-electron temperature ratios, specifically $\tau=100$ and $\tau=1$. We consider these two cases representative of the finite-$\tau$ and large-$\tau$ regimes, respectively. In both  regimes, we observed the existence of strong velocity shears that initiate Kelvin-Helmholtz instabilities. These instabilities lead to the propagation of turbulence through the separatrices and the formation of eddies. We will discuss the nature of this turbulence for both cases.

The paper is organized as follows. In Section \ref{sec:model},  we present the gyrofluid model and the different limiting regimes  that it can cover. In Section \ref{sec:linear}, we investigate the linear growth rates of the tearing instability in different parameter regimes. Section \ref{sec:nonlin} focuses on the turbulence dynamics and the vortex formation that develop at longer times. Section \ref{sec:conclusion} is the Conclusion.

\section{Model equations}\label{sec:model}
We make use of the gyrofluid model consisting of the two evolution equations
\begin{align}
&\frac{\pa N_e}{\pa t}+ [\phi - \rho_s^2 \bpar , N_e ] -[\apar ,\lapp \apar]=0, \label{contfin}\\
&\frac{\pa }{\pa t}(\apar - d_e^2 \lapp \apar)+[\phi - \rho_s^2 \bpar , \apar - d_e^2 \lapp \apar]+\rho_s^2 [\apar , N_e]=0, \label{ohmfin}
\end{align}
complemented by the static relations
\begin{align}
&N_e + (1 - \Gamma_{0i} (b_i) +\Gamma_{1i} (b_i))\bpar = \left(\frac{\Gamma_{0i} (b_i) -1}{\tau}+\frac{\beta_e}{2}d_e^2 \lapp\right)\frac{\phi}{\rho_s^2}, \label{qnfin}\\
& \left( \frac{2}{\beta_e} + (1 + 2 \tau) (\Gamma_{0i} (b_i) -\Gamma_{1i} (b_i))\right) \bpar=\left(1 - \frac{\Gamma_{0i} (b_i) -1}{\tau} -\Gamma_{0i} (b_i) +\Gamma_{1i} (b_i) \right) \frac{\phi}{\rho_s^2}. \label{ampperpfin}
\end{align}
This system is the two-dimensional reduction of a model derived in Ref. \cite{Pas18}. The  latter is formulated in a slab geometry, adopting Cartesian coordinates $\{x,y,z\}$ and assuming the presence of a strong magnetic guide field along the unit vector  ${\widehat{\boldsymbol z}}$. In the 2D version adopted here, we assume that the dynamical variables do not depend on the $z$ coordinate.  

Equations (\ref{contfin}) and (\ref{ohmfin}) correspond to the continuity equation for the electron gyrocenter density fluctuations $N_e$ and to Ohm's law, respectively. The relations (\ref{qnfin}) and (\ref{ampperpfin}), on the other hand, express the quasi-neutrality condition and the perpendicular component of Amp\`ere's law, respectively. Here, $A_\|$ and $B_\|$ are the components, along the guide field, of the magnetic potential and of the magnetic fluctuations, whereas $\phi$ indicates the electrostatic potential. The expression for the total magnetic field is given by
\beq  \label{magfield}
\mathbf{B}(x,y,t)=\nabla \apar (x,y,t)\times {\widehat{\boldsymbol z}}+(1+\bpar (x,y,t)){\widehat{\boldsymbol z}}.
\eeq
In Eqs. (\ref{contfin})-(\ref{ampperpfin}) and \eqref{magfield}, the variables are dimensionless and expressed according to the following normalization:
 \begin{align}
&t=\frac{v_A}{L}\hat{t}, \qquad x=\frac{\hat{x}}{L}, \qquad y=\frac{\hat{y}}{L},  \nno\\
&N_e=\frac{L}{\hat{d}_i}\frac{\hat{N}_e}{n_0}, \qquad \phi=\frac{c}{v_A}\frac{\hat{\phi}}{L B_0}, \qquad \apar=\frac{\hat{A}_\parallel}{L B_0}, \qquad \bpar=\frac{L}{\hat{d}_i}\frac{\hat{B}_\parallel}{B_0}, \label{norm}
\end{align}
where the hat denotes dimensional variables, $L$ is a characteristic scale length, $n_0$ is the equilibrium density, $B_0$ the guide field amplitude, $c$ the speed of light, $v_A=B_0/\sqrt{4 \pi m_i n_0}$ the Alfv\'en speed, with $m_i$ indicating the ion mass, whereas $\hat{d}_i=c/\sqrt{4 \pi n_0 e^2/m_i}$ is the ion skin depth, with $e$ corresponding to the elementary charge. This  normalization differs from that adopted in Ref. \cite{Pas18}. In the present paper, we  opted for the normalization (\ref{norm}) because this will help in establishing contact with previous results present in the literature. Also, the normalization (\ref{norm}) might be more  appropriate for astrophysical applications.

 Parameters of the system are the electron skin depth $d_e=(1/L)c/\sqrt{4 \pi n_0 e^2/m_e}$, with $m_e$ indicating the electron mass, the ratio $\beta_e=8 \pi n_0 T_{e0}/B_0^2$ between electron kinetic and guide-field magnetic pressures, the ion sonic Larmor radius $\rho_s=\sqrt{\beta_e/2}\sqrt{m_i/m_e}d_e$  and the ion-to-electron equilibrium temperature ratio $\tau=T_{i0}/T_{e0}$.
 
 The model equations also involve the canonical bracket $[f,g]=\pa_x f \pa_y g - \pa_y f \pa_x g$ and of the ion FLR operators $\Gamma_{ni} (b_i)$, for $n=0,1$, which correspond, in Fourier space, to multiplication by $I_n (b_i)\exp(-b_i)$, where $I_n$ is the modified Bessel function of first type of order $n$ and $b_i=\tau \rho_s^2 k_\perp^2$, with $k_\perp^2$ indicating the squared modulus of the wave number in the plane perpendicular to the guide field. We also indicated with $\lapp$ the perpendicular Laplacian operator defined by $\lapp f=\pa_{xx} f +\pa_{yy} f$.

 The independent variables of the model vary on a domain $\mathcal{D}=\{(x,y): \, -L_x \leq x \leq L_x \, , \, -L_y \leq y \leq L_y \}$, where $L_x$ and $L_y$ are positive constants. Periodic conditions are imposed at the boundaries of $\mathcal{D}$.
 
 This model can describe the dynamics of collisionless plasmas at low-$\beta_e$, accounting for ion FLR effects as well as electron inertia, which can break the frozen-in condition and allow for magnetic reconnection. The electron fluid is assumed to be isothermal, whereas the fluctuations of the ion gyrocenter moments are neglected in Eqs. (\ref{contfin})-(\ref{ampperpfin}). The model was derived taking
 \beq  \label{delta}
 \delta=\sqrt{\frac{m_e}{m_i}}=\sqrt{\frac{\beta_e}{2}}\frac{d_e}{\rho_s}
 \eeq
 as a small expansion parameter, and  assuming $\beta_e \sim \delta$ for $\delta \rightarrow 0$. We also recall that, although the low-$\beta_e$ limit tends to suppress electron FLR effects, our model retains one contribution which descends from an electron FLR term present in the parent gyrofluid model \cite{Bri92}. This corresponds to the last term in Eq. (\ref{qnfin}),  which becomes relevant in the  large-$\tau$ limit, where it gets comparable to the retained contributions at scale $d_e$. 

 We point out that, in the 2D version that we adopt here, the conservation laws differ considerably from those of the 3D version of the model. Indeed, as is typically the case with Hamiltonian reduced fluid models, in the 2D limit, the system acquires an infinity of Casimir invariants. For Eqs. (\ref{contfin})-(\ref{ampperpfin}), these correspond to the two infinite families
 \beq
 \mathcal{C}_\pm=\int d^2 x \, C_{\pm} (\apar - d_e^2 \lapp \apar \pm d_e \rho_s N_e),
 \eeq
 where $C_{\pm}$ are arbitrary functions. All such functionals are conserved and constrain the dissipationless dynamics. The above infinite families include also quadratic functionals such as those used, for instance in Ref. \cite{Cam96}, to investigate 2D drift-wave turbulence. Namely from a linear combination of quadratic Casimir invariants \cite{Pas18}, one obtains in particular the generalized cross-helicity 
 \beq  \label{crosshel}
 \mathcal{H}_C=\int d^2 x \, N_e (\apar - d_e^2 \lapp \apar),
 \eeq
 which is then also a conserved quantity for our 2D model.
 
 In the general 3D case, on the other hand, instead of the infinite families $\mathcal{C}_\pm$, one has the two linear Casimir invariants
 \begin{align}
& \mathcal{C}_1=\int d^3 x \, (\apar - d_e^2 \lapp \apar + d_e \rho_s N_e),\\
&\mathcal{C}_2=\int d^3 x \, (\apar - d_e^2 \lapp \apar - d_e \rho_s N_e).
 \end{align}
 A generalized cross-helicity (with an expression analogous to Eq. (\ref{crosshel})) is also conserved in 3D, but it is no longer a linear combination of Casimir invariants.

 As already pointed out in Refs. \cite{Pas18,Pas19}, Eqs. (\ref{contfin})-(\ref{ampperpfin}) generalize reduced models previously presented in the literature and which can be retrieved in the appropriate limits. In the following, we briefly review the different limits that will be relevant for the subsequent analysis.
Specifically, under the conditions of finite $k_\perp \rho_s$ and $\beta_e \ll 1$, the three regimes in $\tau$ are described in Subsections A, B, and C. These subsections correspond to scenarios where $k_\perp \rho_i$ takes values of significantly less than 1, around 1, and greater than 1, respectively.

 \subsection{Cold-ion limit: $\tau \ll 1$}
 
 In this limit, the model (\ref{contfin})-(\ref{ampperpfin}) reduces to
\begin{align}
& \frac{\pa \lapp \phi}{\pa t}+[\phi , \lapp \phi] - [\apar , \lapp \apar]=0,  \label{contci}\\
&\frac{\pa }{\pa t}(\apar - d_e^2 \lapp \apar)+[\phi, \apar - d_e^2 \lapp \apar]+{\rho_s'}^2[\apar, \lapp \phi]=0,  \label{ohmci}\\
&N_e=\lapp \phi,  \label{neci}\\
&\bpar = -\frac{\beta_e}{2 + \beta_e} \lapp \phi,  \label{bparci}
\end{align}
where we have neglected contributions $\delta^2$ times smaller than the leading order terms in each evolution equation, assuming $\tau=\mathcal{O}(\delta^2)$ and $\beta_e/2 \sim d_e^2 \sim \delta$. In Eq. (\ref{ohmci}) we introduced the parameter $\rho_s' = \rho_s \left(1 - {\beta_e}/4\right)$, which is a modified ion sonic Larmor radius, accounting for parallel magnetic fluctuations effects. If $\beta_e$ is assumed even smaller, say $\beta_e = \mathcal{O}(\delta^2)$, then Eqs. (\ref{contci})-(\ref{ohmci}) identify with the two-field reduction, adopted for instance in Refs. \cite{Caf98,Gra01}, of the three-field model of Ref. \cite{Sch94}. Note that, in this limit, in order to satisfy the relation (\ref{delta}), one also needs $\rho_s \rightarrow 0$ as $\delta \rightarrow 0$.

 \subsection{Finite $\tau$ and negligible $\beta_e$}

Here,  contributions of order $\beta_e$ are assumed negligible in the model equations. Performing a Pad\'e approximation of the operator $\Gamma_{0i}$ (as done in Refs. \cite{Gra00,Del11}) one obtains, from Eq. (\ref{qnfin}), the relation $\phi=\lappinv N_e -\tau \rho_s^2 N_e$, which, inserted into Eqs. (\ref{contfin})-(\ref{ohmfin}), yields
\begin{align}
&\frac{\pa N_e}{\pa t}+[\lappinv N_e , N_e] - [\apar , \lapp \apar]=0,  \label{conttau2}\\
&\frac{\pa }{\pa t}(\apar - d_e^2 \lapp \apar)+[\lappinv N_e, \apar - d_e^2 \lapp \apar]+{\rho_\tau}^2[\apar, N_e]+d_e^2 \tau \rho_s^2 [N_e , \lapp \apar]=0,  \label{ohmtau2}
\end{align}
where
\beq
\rho_\tau=\sqrt{1+\tau}\rho_s.
\eeq
 This model was  used in Refs. \cite{Gra00,Gra10,Del11} to study ion FLR effects on collisionless magnetic reconnection.
 
 \subsection{Hot-ion limit: $\tau \gg 1$}
 \label{ssec:hotion}
 
 If we take $\tau \sim 1/\delta^2$, with $\beta_e/2 \sim d_e^2 \sim \delta$ and again neglect contributions $\mathcal{O}(\delta^2)$ times smaller than the dominant terms in each evolution equation, we obtain
 \begin{align}
&\frac{\pa}{\pa t}\left(d_e^2 \lapp - \left(\frac{2}{\tau \beta_e}+1\right)\right)\frac{\phi}{\rho_s^2}+\left[\phi, d_e^2 \lapp \frac{\phi}{\rho_s^2}\right]-\frac{2}{\beta_e}[\apar, \lapp \apar]=0,  \label{conthi}\\
&\frac{\pa }{\pa t}(\apar - d_e^2 \lapp \apar)+[\phi, \apar - d_e^2 \lapp \apar]=0, \label{ohmhi} \\
&N_e=\frac{\beta_e}{2}\left(d_e^2 \lapp -\left(\frac{2}{\tau \beta_e}+1\right)\right)\frac{\phi}{\rho_s^2},  \label{qnhi}\\
&\bpar=\frac{\beta_e}{2}\frac{\phi}{\rho_s^2},  \label{ampperphi}
\end{align}
 which corresponds, up to the normalization, to the model for IKAW turbulence introduced in Ref. \cite{Che17, Pas17}. 
 
 If we increase $\tau$ even further (say $\tau = \mathcal{O}(1/\delta^3)$), so that the contribution of $2/(\tau \beta_e)$ becomes negligible with respect to $1$, the evolution equations of the model reduce to
 \begin{align} 
&\frac{\pa}{\pa t}\left(d_e^2 \lapp -1\right)\frac{\phi}{\rho_s^2}+\left[\phi, d_e^2 \lapp \frac{\phi}{\rho_s^2}\right]-\frac{2}{\beta_e}[\apar, \lapp \apar]=0,  \label{contemhd1}\\
&\frac{\pa }{\pa t}(\apar - d_e^2 \lapp \apar)+[\phi, \apar - d_e^2 \lapp \apar]=0. \label{ohmemhd1}
\end{align}
Making use of Eq. (\ref{ampperphi}) and introducing the whistler time $t_w=d_i t$, Eqs. (\ref{contemhd1})-(\ref{ohmemhd1}) can be rewritten as
\begin{align}
&\frac{\pa}{\pa t_w}\left(d_e^2 \lapp - 1\right)b+[b, d_e^2 \lapp b]-[\apar, \lapp \apar]=0,  \label{contemhd2}\\
&\frac{\pa }{\pa t_w}(\apar - d_e^2 \lapp \apar)+[b, \apar - d_e^2 \lapp \apar]=0, \label{ohmemhd2} 
\end{align}
where $b=d_i \bpar$. Equations (\ref{contemhd2})-(\ref{ohmemhd2}) correspond to the equations of 2D REMHD, with electron inertia which, in two dimensions, are formally identical to those of 2D incompressible EMHD (see, e.g., Refs. \cite{Kin90,Bis00}).
We will therefore refer to the above limit as to the REMHD limit. However, later on, since the 2D equations are identical to those of EMHD, we will compare the numerical growth rates of the tearing modes to the theoretical formulas obtained within the framework of EMHD. 
We remark that, in the 3D case, the presence of the strong guide field ordering, would lead to a model for Inertial Whistler Turbulence \cite{Che17}. We also point out that the above procedure leading to REMHD upon neglecting corrections of order $\delta^2$ makes it possible to retain corrections of order $\delta$, such as those that become relevant at the scale $d_e$. At the same time, it discards contributions at the scale of the electron Larmor radius $\rho_e = \delta \rho_s$.  Indeed, even if the only electron FLR term retained in the model leads to a contribution in Eq. (\ref{contfin}) that is comparable to the other terms at scale $d_e$ when $\tau\gg 1$, it also leads in Eq. (\ref{ohmfin}) to terms that become relevant at scales of order  $\rho_e $. Since other electron FLR contributions are missing in the model, the description at the scale $\rho_e$ is not accurate. Therefore, it is necessary to ensure that the length scales involved in the solution remain significantly larger than the electron Larmor radius during the evolution. This issue will be further discussed in Sec. \ref{sssec:emhdlin} in the context of tearing modes.

\section{Linear tearing modes}\label{sec:linear}

In this Section,  combining numerical and analytical studies, we analyze how the variation of the ion-to-electron temperature ratio $\tau$ affects the linear growth rate of tearing modes. We  consider both the so-called "constant-$\psi$" regime \cite{Fur63}, typically valid for small values of $\Delta'$, and the regime with large $\Delta'$. We recall that the parameter $\Delta'$ \cite{Fur63} is the variation, at the resonant surface, of the logarithmic derivative of the amplitude of outer solution in the linearized model equations and corresponds to the standard stability parameter of tearing modes.

In the following, we begin by describing the adopted equilibrium state, and subsequently treat the constant-$\psi$ and large $\Delta '$ cases.

\subsection{Equilibrium state}

 We linearize the model equations (\ref{contfin})-(\ref{ohmfin}) about the equilibrium solution considered in Ref. \cite{Porcelli02}
\beq  \label{equilgf}
\phi^{(0)} = 0, \qquad \apar^{(0)} (x)= \frac{\ba}{2\cosh^2(x)}, 
\eeq
where  $\ba$ is a constant determining the magnetic equilibrium amplitude.  The corresponding poloidal equilibrium magnetic field is given by
\beq  \label{mageq}
\mathbf{B}_{eq \perp}(x)=\nabla \apar^{(0)} (x)\times {\widehat{\boldsymbol z}} =\ba \frac{\sinh(x)}{\cosh^3 (x)} {\widehat{\boldsymbol y}},
\eeq
where ${\widehat{\boldsymbol y}}$ is the unit vector along the $y$ coordinate. We set $\ba=2.598$, so that, for $L_x > 0.6586$, one has $\max\limits_{-L_x \leq x \leq L_x} \vert \mathbf{B}_{eq \perp} (x)\vert =1$. Note also that, from  Eqs. (\ref{qnfin})-(\ref{ampperpfin}), the choice $\phi^{(0)} = 0$ yields $N_e^{(0)}=\bpar^{(0)}=0$ at equilibrium. The choice of the equilibrium profile is motivated by the rapid decay to zero of this profile and its derivatives for $x \to \pm\infty$. This allows for the imposition of periodic boundary conditions, given a sufficiently large integration domain. This setting enables us to use a Fourier spectral code, most suitable for simulating gyrofluid models with nonlocal operators $\Gamma_0$ and $\Gamma_1$, taking the form of Fourier multipliers, as demonstrated in \cite{Gra07}.

This equilibrium corresponds to a current sheet centered at $x=0$, with a dimensionless length of $2L_y=2\hat{L}_y/L$, and a dimensionless width corresponding to unity. In other words, one uses the typical width of the current sheet as length unit. In our numerical simulations, the perturbation of the equilibrium potential $\apar^{(0)}$ is a function $\apar^{(1)}(x,y,t)$, whose dependence on $y$ is of the form $\apar^{(1)} \propto \cos (k_y y)$, with $k_y = \pi m/L_y $ (where $m$ is an integer) and is initially excited by the mode $m=1$. The stability condition is given by the tearing parameter, which for  equilibrium \eqref{equilgf} is (see \cite{Porcelli02})
\begin{equation}  \label{Deltaprime}
\Delta'= 2 \frac{\left(5- k_y^2\right) \left( k_y^2+3\right)}{ k_y^2 ( k_y^2+4)^{1/2}}.
\end{equation}
The equilibrium (\ref{equilgf}) is unstable to tearing when $\Delta'(k_y) >0$, corresponding to wave numbers $k_y  < \sqrt{5}$. In limiting cases, such as $\tau \ll 1$ and $\tau \gg 1$, where the gyroaverage operators become trivial, and the model equations become local in the space variables, other equilibrium profiles could be considered, such as the Harris sheet \cite{Bir01} or a sinusoidal profile \cite{Ott93}. Such equilibria have a different expressions for $\Delta'$ and therefore the corresponding linear growth rates of the tearing instability display a different dependence on $k_y$.

The linear and maximum growth rate of the tearing instability are measured at the X-point using the following quantity:
\begin{equation}
\gamma = \frac{d}{dt} \ln\left| \apar^{(1)} \left(0,0,t \right) \right|,
\label{numericalgrowthrate}
\end{equation}
which is constant in time during the linear phase.

\subsection{"Constant-$\psi$" regime}   \label{ssec:cpsi}

In the "constant-$\psi$" regime, it is assumed that the amplitude of the perturbation $\apar^{(1)}$ is approximately constant in the inner region centered around the resonant surface. This assumption is valid when $\Delta ' \epsilon \ll 1$, where $\epsilon$ denotes the width of the inner region, which  is estimated below.

We first present analytical results concerning the cold-ion and EMHD regimes, that will then be compared  with the numerical values obtained by the simulations in the appropriate limits.

\subsubsection{Dispersion relation for the cold-ion regime}  \label{sssec:cilin}

We consider Eqs. (\ref{contci})-(\ref{ohmci}) and their linearization about the equilibrium
(\ref{equilgf}). We assume perturbations of the form 
\beq  \label{perturb}
\apar^{(1)}(x,y,t)=\tilde{A}(x)\exp(ik_y y+\gamma t) + \mathrm{c. c.}, \qquad \phi^{(1)}(x,y,t)=\tilde{\phi}(x)\exp(ik_y y+\gamma t) + \mathrm{c. c.},
\eeq
where $\mathrm{c. c.}$ indicates complex conjugate. In Eq. (\ref{perturb}), the constant $\gamma$  provides the growth rate of the tearing perturbation. We assume that the amplitudes $\tilde{A}$ and $\tilde{\phi}$ are even and odd functions of $x$, respectively, which is the standard parity for tearing modes. 

When $\beta_e$ is neglected compared to unity, it is known that the dispersion relation for tearing modes for  the considered equilibrium is given by \cite{Por91,Bet22}
\beq  \label{disprelci}
\gamma= \ba k_y d_e \rho_s \frac{\Delta'}{\pi},
\eeq
with $\Delta'$ given by Eq. (\ref{Deltaprime}).

We remark that small corrections in $\beta_e$ due to parallel magnetic perturbations only affect Eq. (\ref{ohmci}), in particular by turning $\rho_s^2$ into ${\rho_s'}^2$. Therefore, such effects can be accounted for, in the dispersion relation, in a straightforward way, by replacing $\rho_s$ with $\rho_s'$ in Eq. (\ref{disprelci}). Because $\rho_s' \leq \rho_s$, we can infer that parallel magnetic perturbations tend to decrease the growth rate of tearing modes.  We recall that a similar result was obtained in Ref. \cite{Gra22}. In this Reference, the combined effects of parallel magnetic perturbations and electron FLR terms were considered, but the latter ones were not treated self-consistently, although the corresponding relation showed good agreement with numerical results. The present model neglects electron FLR terms in the cold-ion limit but retains small corrections due to parallel magnetic perturbations. Therefore, it  makes it possible  to isolate and identify, in a more rigorous way, the stabilizing effect of a finite but small $\bpar$.

\subsubsection{Dispersion relation for the EMHD regime} \label{sssec:emhdlin}
For the EMHD model in the constant-$\psi$ regime, an analytical dispersion relation for tearing modes is given in Ref. \cite{Bul92}. A thorough investigation of EMHD tearing modes, including a numerical validation of the dispersion relation of Ref. \cite{Bul92}, was recently presented in Ref. \cite{Bet23}. However, applying the result of Ref. \cite{Bul92} to our gyrofluid model in the large $\tau$ limit is more delicate.  As anticipated in Sec. \ref{ssec:hotion}, the cold-ion and EMHD systems, corresponding to Eqs. (\ref{contci})-(\ref{ohmci}) and (up to the normalization) (\ref{contemhd1})-(\ref{ohmemhd1}), respectively, were obtained from the original gyrofluid model, assuming that length scales remain of the order of the characteristic scale length $L$. The latter, recalling that $x=\hat{x}/L$ and considering Eq. (\ref{mageq}), can be considered as a characteristic scale of variation of the equilibrium magnetic field. However, in the tearing mode analysis, an inner region, involving small scales, is present around the resonant surface,  where terms with high order derivatives, that are negligible on scales of order $L$, can become relevant. In the cold-ion case, one can safely insert the relations (\ref{neci})-(\ref{bparci}) into Eqs. (\ref{contfin})-(\ref{ohmfin}), neglect corrections which are $\delta^2$ smaller and thus obtain Eqs. (\ref{contci})-(\ref{ohmci}), for which the dispersion relation (\ref{disprelci}) is valid. Indeed, it thus turns out that terms which are $\delta^2$ smaller than the leading terms, are negligible both in the inner region and in the remaining outer region, without imposing further conditions between the growth rate and the parameters $\beta_e, d_e $ and $\rho_s$. However, for the hot-ion case, in general, this is not the case. Therefore, we first examine the conditions under which the EMHD tearing mode relation of Ref. \cite{Bul92} can be applied to the hot-ion limit of our gyrofluid model.

Because we are interested in the EMHD limit, rather than in the IKAW limit, we assume that $\tau$ is large enough, so that $2/(\tau \beta_e)$ is neglected compared to $1$ in Eq. (\ref{qnhi}). Then we insert the resulting hot-ion relations
\begin{align}
&N_e=\frac{\beta_e}{2}\left(d_e^2 \lapp -1\right)\frac{\phi}{\rho_s^2},  \label{qnhie}\\
&\bpar=\frac{\beta_e}{2}\frac{\phi}{\rho_s^2},  \label{ampperphie}
\end{align}
 into Eqs. (\ref{contfin})-(\ref{ohmfin}) but {\it without } neglecting any terms, so as to account for the possibility that terms negligible on scales of order $L$ might become non-negligible in the inner region.
 Upon linearizing the resulting system about the equilibrium (\ref{equilgf}), we obtain
 \begin{align}
 &\frac{\beta_e}{2 \rho_s^2}\frac{\gamma}{k_y}\left[ d_e^2 (\tphi '' -k_y^2 \tphi) -\tphi\right]-i\left[{\aeq} ' (\tapar '' - k_y^2 \tapar)-{\aeq} ''' \tapar \right]=0, \label{linhi1}\\
 &\frac{\gamma}{k_y}\left[ \tapar -d_e^2 (\tapar '' - k_y^2 \tapar)\right]-i {\aeq}' \tphi + i d_e^2 {\aeq} ''' \tphi -i \frac{\beta_e}{2}d_e^2  {\aeq} ''' \tphi + i \frac{\beta_e}{2} d_e^2 {\aeq} ' (\tphi '' -k_y^2 \tphi)=0, \label{linhi2}
 \end{align}
where the prime denotes derivative with respect to the argument of the function, which is $x$ in this case. We consider the system (\ref{linhi1})-(\ref{linhi2}) on an infinite domain (which is a reasonable approximation if $L_x$ is sufficiently large) with $\tapar \rightarrow 0 $ and $\tphi \rightarrow 0$ for $x \rightarrow \pm \infty$.

The EMHD case is retrieved from Eqs. (\ref{linhi1})-(\ref{linhi2}) if the last three terms on the left-hand side of Eq. (\ref{linhi2}) are neglected and the change of variables $\gamma= d_i \gamma_w, \, \tphi = d_i \tilde{b}$ is performed. In the latter expressions we indicated with $\gamma_w$ the growth rate in terms of whistler units of time and with $\tilde{b}$ the amplitude of the perturbed parallel magnetic field. 
%These three terms are all proportional to $\beta_e$, but $\beta_e$ is also present in Eq. (\ref{linhi1}), which must not change to retrieve EMHD, so one cannot simply set $\beta_e=0$. 

Assuming $\gamma \ll 1$ and $d_e \ll 1$, in the outer region, far from the resonant surface at $x=0$, all terms proportional to $d_e^2$ and $\gamma$ are negligible. These include also the three above mentioned "non-EMHD" terms, which confirms that, on scales of the order of the equilibrium magnetic field, the tearing analysis of the system (\ref{linhi1})-(\ref{linhi2}) coincides with that of EMHD. However, around $x=0$, where ${\aeq}'$ and ${\aeq} '''$ vanish, terms negligible in the outer region might become important. To see at which scale such terms become non-negligible one introduces rescaled variables
\beq
X=\frac{x}{\epsilon}, \qquad \tapar_{in}(X)=\tapar (x), \qquad \txi (X)=-\frac{i}{g}\tphi (x),
\eeq
where 
\beq
\epsilon \ll 1
\eeq
is a stretching factor and 
\beq
g=\frac{\gamma}{k_y}.
\eeq
Given that $\epsilon \ll 1$, around $x=0$ the system (\ref{linhi1})-(\ref{linhi2}) can be approximated by
\begin{align}
&\frac{\beta_e}{2 \rho_s^2} g^2 \left( \frac{d_e^2}{\epsilon^2} \frac{d^2 \txi}{dX^2} - \txi\right)+\ba \frac{X}{\epsilon} \frac{d^2 \tapar_{in}}{dX^2}=0,  \label{innhi1}\\
&1 - \frac{d_e^2}{\epsilon^2}\frac{d^2 \tapar_{in}}{dX^2}-\ba \epsilon X \txi +\frac{\beta_e}{2} d_e^2 \ba \frac{X}{\epsilon} \frac{d^2 \txi}{dX^2}=0,  \label{innhi2}
\end{align}
where we made use of the constant-$\psi$ approximation and set $\tapar_{in} \approx 1$. Two out of the three non-EMHD terms turned out to be negligible also in the inner region. However, a term remains, which is given by the last term on the left-hand side of Eq. (\ref{innhi2}). This term originates from the electron FLR effect retained in Eq. (\ref{qnfin}) and is negligible, compared to the third term on the left-hand side of Eq. (\ref{innhi2}), if
\beq  \label{condemhd}
\epsilon \gg \sqrt{\frac{\beta_e}{2}}d_e=\rho_e,
\eeq
where $\rho_e $ is the thermal electron Larmor radius. Therefore, the EMHD case is retrieved if the width of the inner region is much larger than the electron Larmor radius.

Note that, in the cold-ion case, retaining the electron FLR correction, from Eq. (\ref{qnfin}), leads to $N_e=(1 + \delta^2) \lapp \phi$. Therefore, the electron FLR term, in this case, corresponds to a small correction of order $\delta^2$ to the ion polarization term, and can thus be safely neglected, which leads to the relation (\ref{neci}). However, the requirement that the width of the inner region should be much larger than the thermal electron Larmor radius, should hold also in the cold-ion limit, as pointed out in Refs. \cite{Fit10,Gra22}. 

The condition (\ref{condemhd}) can be made more explicit once one finds the expression for $\epsilon$ in terms of the parameters of the system. This corresponds to the distinguished limit of the system. Assuming that the condition (\ref{condemhd}) holds, from Eqs. (\ref{innhi1})-(\ref{innhi2}) one obtains
\beq   \label{eqinn}
\frac{d^2 \bxi}{dX^2}-\frac{\epsilon^2}{d_e^2}\bxi-\frac{\epsilon^4}{\delta^2 d_e^2 g^2}{\ba}^2 X^2 \bxi+ \ba X=0,
\eeq
where $\bxi=(\delta^2 d_e^2 g^2/\epsilon^3)\txi$.

The second term on the left-hand side of Eq. (\ref{eqinn}) would be of the same order of the first term, if $\epsilon$ were equal to $d_e$. However, this would make the third term diverge, because its coefficient would become $d_i^2 \ba^2 X^2/g^2$, which goes to infinity as $g \rightarrow 0$. Consequently, the second term has to be subdominant. We remark that, if we had kept the contribution proportional to $2/(\tau \beta_e)$ of the hot-ion limit, the second term on the left-hand side of Eq. (\ref{eqinn}) would have been multiplied by a factor $(1+2/(\tau \beta_e))$. If this factor remains of order unity, that term would have still been subdominant.  Therefore, the presence of the coefficient characteristic of  the IKAW regime does not affect  the tearing dispersion relation, which is  consistent with the analysis of Ref. \cite{Bol19}. 

The distinguished limit is then provided by balancing the first and the third term, which leads to
\beq
\epsilon=\left(\frac{\delta d_e g}{\ba}\right)^{1/2}.
\eeq

The condition (\ref{condemhd}) can then be reformulated, in terms of the original parameters of the system, as
\beq  \label{condemhd2}
\gamma \gg k_y \sqrt{\frac{\beta_e}{2}} \rho_s.
\eeq
If the condition (\ref{condemhd2}) is fulfilled, the tearing analysis of Ref. \cite{Bul92} applies to the hot-ion limit of our gyrofluid model, which yields
\beq  \label{disprelemhd}
\gamma= d_i \ba k_y \left(\frac{\Gamma (1/4)}{2 \pi \Gamma(3/4)}\right)^2 d_e^2 {\Delta '}^2,
\eeq
where $\Gamma (x)$ is the Gamma function. We also remark that, with respect to the original Ref.\cite{Bul92}, the expression (\ref{disprelemhd}) contains a factor $d_i$, due to the fact that $\gamma$ is normalized in terms of Alfv\'en units of time.

Assuming the validity of the relation (\ref{disprelemhd}) and inserting it into Eq. (\ref{condemhd2}), one can also re-express the latter condition as
\beq   \label{condemhd3}
\frac{\beta_e}{2}\ll 0.22 \ba d_e^2 {\Delta '}^2.
\eeq
In this formulation, the condition for excluding the electron FLR contribution only depends on the plasma parameters $\beta_e$ and $d_e$ and on the parameters $\ba$ and $\Delta '$ associated with the equilibrium and the wave number of its perturbation.

\subsubsection{Numerical results and comparison with analytical dispersion relations}

Here we analyze, by means of numerical simulations, the tearing growth rates for different values of the parameters and compare the results, in the appropriate limits, with the analytical predictions of 
Secs.  \ref{sssec:cilin} and \ref{sssec:emhdlin}. 
Equations (\ref{contfin}) - (\ref{ampperpfin}) are solved with a Fourier pseudo-spectral solver on a periodic domain, with advancement in time performed by a third-order Runge-Kutta numerical scheme.
To investigate the linear growth rate, we consider the case of small $\Delta'$, where the "constant-$\psi$" approximation is applicable. In this context, the initial wave number is set to $k_y = 1.923$, leading to $\Delta' = 1.699$.

To study the  influence of $\tau$, we conducted a series of numerical simulations covering a range of equilibrium temperature ratios spanning from $\tau=0.1$  to $\tau=1000$. In terms of the ion thermal Larmor radius $\rho_i=\sqrt{\tau} \rho_s$, the range corresponds to going from  $\rho_i=0.09$ to $\rho_i=9.48$. For these simulations we fixed the sound Larmor radius  to $\rho_s = 0.3$ and the mass ratio to ${m_e}/{m_i} = 2 \times 10^{-5}$. The use of such a small mass ratio, which is made possible by the computational efficiency of the gyrofluid modeling,  ensures a small electron skin depth and a small $\beta_e$, meeting the validity criteria for the linear theory.

\begin{figure}
    \centering
\includegraphics[trim=0 0 0 0, scale=0.7]{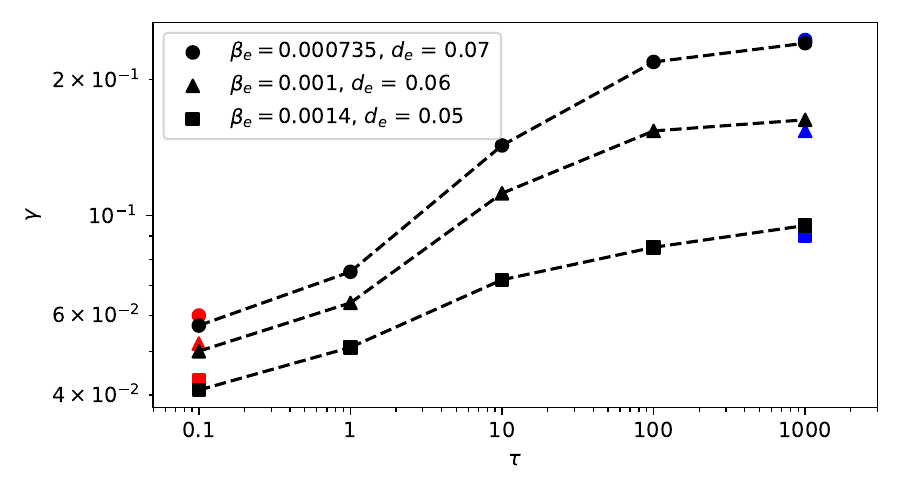}
 \caption{ Linear growth rate as a function of $\tau$ for different values of $\beta_e$ and $d_e$. The normalized ion Larmor radius becomes larger than unity starting from $\tau = 10$. Red symbols indicate the theoretical predictions based on the relation (\ref{disprelci}) valid in the cold-ion case. Blue symbols represent the theoretical predictions based on Eq. (\ref{disprelemhd}) and corresponding to  the EMHD regime.}
    \label{fig:cstpsi} 
\end{figure}

Figure \ref{fig:cstpsi}  displays the linear growth rates for three different values of $\beta_e$ and $d_e$ given by  simulations together with the asymptotic values predicted by the theory. The range of parameters has been chosen in such a way that, for $\tau \gg 1$,  condition (\ref{condemhd2}) is satisfied, so that the EMHD regime should be recovered. For $\tau = 0.1$, the numerical growth rates closely align with those predicted by the cold-ion relation (\ref{disprelci}), shown in red on the figure (the adopted values of $\beta_e$ in this case are so low that the use of the effective parameter $\rho_s'$, mentioned in Sec. \ref{sssec:cilin}, is practically of no use).
As we increase $\tau$, the growth rates also increase and the model converges towards the EMHD model described in Sec. \ref{ssec:hotion}. For $\tau = 1000$, the linear growth rates match the analytical estimates based on Eq. (\ref{disprelemhd}) and represented in blue on the figure.

Although the values of $\tau$ of practical physical interest, for instance for space plasmas, concern only a small portion around $\tau \approx 1$ of the considered range, our analysis shows how, by increasing $\tau$, the linearized system can perform a transition from the cold-ion regime, to a regime where the thermal ion Larmor radius is much larger than the characteristic scale of the current sheet, so that the observed dynamics is essentially due to electrons. In this regime the predictions of EMHD apply. In particular,  we  note the transition from a growth rate linear in $d_e \Delta '$, for cold-ions, to a growth rate proportional to $d_i d_e^2 {\Delta '}^2$ for very hot ions.

In Tables \ref{table:EMHD} and \ref{table:Prcelli}, we present a precise quantitative comparison indicating the errors between numerical and theoretical values. Table \ref{table:EMHD} shows the results obtained for $\tau=1000$ compared to the EMHD theoretical formula (\ref{disprelemhd}), while Table \ref{table:Prcelli} shows the numerical results for $\tau=0.1$ compared to the theoretical predictions based on Eq. (\ref{disprelci}). The values of the growth rate  displayed in the Tables correspond mostly to those shown also in Fig. \ref{fig:cstpsi}. The applicability conditions of relations (\ref{disprelci}) and (\ref{disprelemhd}) appear to considerably limit the admissible range of the plasma parameters, if one requires a relative   discrepancy of a few percent between numerical and analytical predictions. However,  we also observe a satisfactory  agreement with the EMHD relation (\ref{disprelemhd})  in the case  $d_e=0.21$ and $\beta_e = 2 \times 10^{-5}$, well different from those shown in Fig. \ref{fig:cstpsi}.

\begin{table}[ht]
    \centering
    \caption{ Comparison, for $\tau=1000$, of  the growth rate  $\gamma$ obtained numerically and  $\gamma_T$, based on Eq. (\ref{disprelemhd}), for different values of the parameters. Relative errors are also indicated.}
    
    \begin{ruledtabular}   
    \begin{tabular}{cccccc}
        \toprule
        $d_e$ & $\beta_e$ & $\gamma$ & $\gamma_T$ & Error (\%) \\
        \midrule
        0.05 & $1.4 \times 10^{-3}$ & $9.3 \times 10^{-2}$ & $8.9 \times 10^{-2}$ & 4.5 \\
        0.06 & $1 \times 10^{-3}$ & $1.63 \times 10^{-1}$ & $1.54 \times 10^{-1}$ & 5.8 \\
        0.07 & $7.3 \times 10^{-4}$ & $2.41 \times 10^{-1}$ & $2.45 \times 10^{-1}$ & 1.6 \\
        0.21 & $2 \times 10^{-5}$ & $1.45 \times 10^{-3}$ & $1.36 \times 10^{-3}$ & 6.2 \\
        \bottomrule
    \end{tabular}
    \end{ruledtabular}  
    
    \label{table:EMHD}
\end{table}

\begin{table}[ht]
    \centering
    \caption{Comparison, for $\tau=0.1$, of  the growth rate  $\gamma$ obtained numerically and  $\gamma_T$, based on Eq. (\ref{disprelci}), for different values of the parameters. Relative errors are also indicated.}
    
    \begin{ruledtabular}   %pls
    \begin{tabular}{cccccc}
        \toprule
        $d_e$ & $\beta_e$ & $\gamma$ & $\gamma_T$ & Error (\%) \\
        \midrule
        0.05 & $1.4 \times 10^{-3}$ & $4.1 \times 10^{-2}$ & $4.4 \times 10^{-2}$ & 6.82 \\
        0.06 & $1 \times 10^{-3}$ & $5 \times 10^{-2}$ & $5.3 \times 10^{-2}$ & 5.66 \\
        0.07 & $7.3 \times 10^{-4}$ & $5.7 \times 10^{-2}$ & $6.2 \times 10^{-2}$ & 8.06 \\
        \bottomrule
    \end{tabular}
    \end{ruledtabular}   %pls
    \label{table:Prcelli}
\end{table}

\begin{figure}
    \centering
\includegraphics[trim=0 0 0 0, scale=0.7]{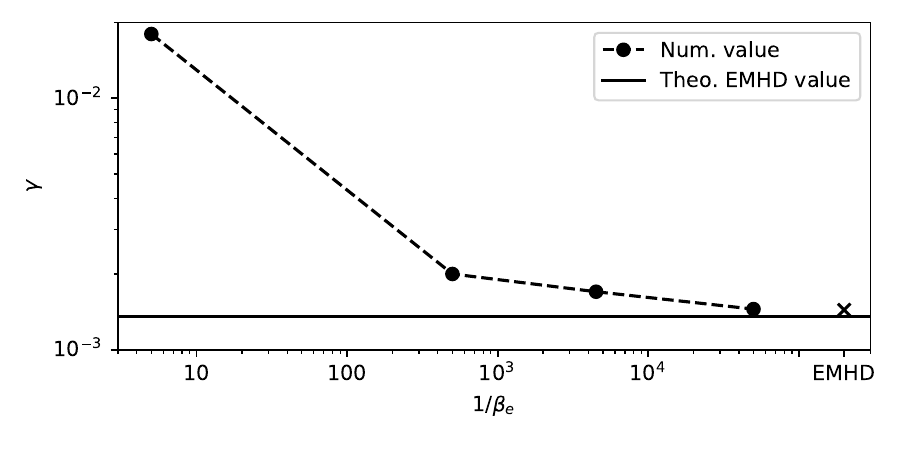}
 \caption{ Linear growth rate as a function of $1/\beta_e$ for $\tau=1000$. As $\beta_e$ decreases, the numerical growth rate approaches the  EMHD analytical prediction \eqref{disprelemhd}   (solid line), also reproduced by direct simulation of the EMHD equations (x-cross). This reflects the fact that one is approaching the asymptotic regime of the condition (\ref{condemhdfig}).}
    \label{fig:cstpsi_2} 
\end{figure}

We also verified how the agreement between the numerical growth rate and the EMHD growth rate obtained from Eq. (\ref{disprelemhd}) improves as the condition (\ref{condemhd2}) is better and better fulfilled in an asymptotic sense.
Figure \ref{fig:cstpsi_2} provides the evolution of $\gamma$ as a function of $1/\beta_e$. In this case, we maintain $\tau = 1000$, and $d_e = 0.21$, and we vary $\beta_e$ and $\rho_s$, while keeping $d_e$ and $d_i$ constant. If one assumes the expression (\ref{disprelemhd}) for the growth rate, the condition (\ref{condemhd2}) becomes
\beq  \label{condemhdfig}
\left(\frac{\Gamma (1/4)}{2 \pi \Gamma(3/4)}\right)^2 d_e^2 {\Delta '}^2 \gg \frac{\beta_e}{2}.
\eeq
Given that we keep $d_e$ and $\Delta '$ fixed, this condition is better and better fulfilled as $\beta_e$ decreases.
The plot includes a theoretical reference line representing the growth rate value derived from the EMHD theory. Interestingly, as $\beta_e$ decreases, corresponding to smaller $\rho_s$ values, our model exhibits a closer agreement with the EMHD results.  The deviations from the theoretical EMHD value as $\beta_e$ increases clearly demonstrate the significant impact of $\beta_e$.

\subsection{Large $\Delta '$ regime}

\begin{figure}
    \centering\includegraphics[trim=0 0 0 0, scale=0.7]{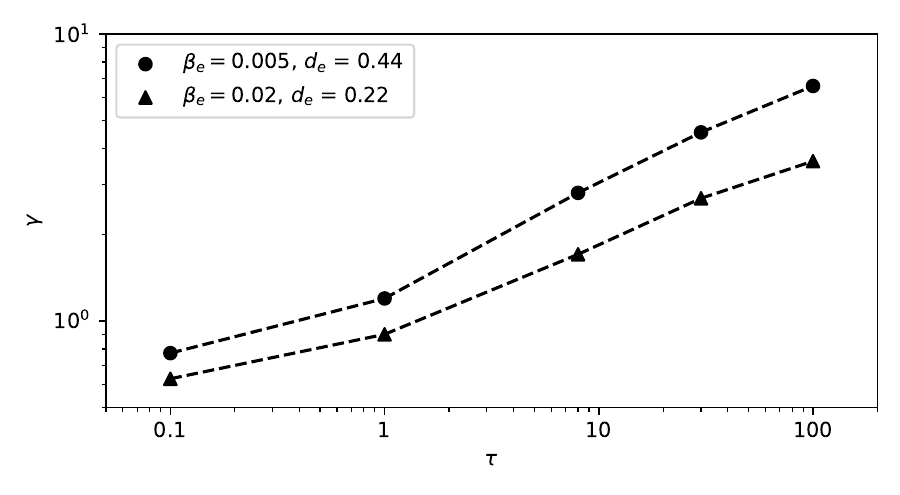}
 \caption{Linear growth rate as a function of $\tau$ for different values of $\beta_e$ and $d_e$, when $\Delta ' = 48.45 $.}    \label{fig:largedeltaprime} 
\end{figure}

We now focus on simulations initiated with a wavenumber of $k_y = 0.55$, yielding $\Delta' = 48.45$.  The ion-sound Larmor radius is fixed to $\rho_s=1$. On Fig. \ref{fig:largedeltaprime} we report the linear growth rate as a function of $\tau$ for both $\beta_e = 0.005$ and $\beta_e = 0.02$. 

The nonlinear evolution of two of these runs, both with $\beta_e=0.02$, is discussed in detail in the next section.

The evolution of the growth rate follows a trend similar to that observed in the case of small $\Delta'$, with $\gamma$ increasing as $\tau$ increases. 
Nevertheless, some differences can be identified. For $\tau=0.1$, the scaling with respect to $d_e$ is rather close to the scaling $\gamma\propto d_e^{1/3}$ predicted in Ref. \cite{Por91}. Indeed, the ratio between the growth rates for $d_e=0.44$ and $d_e=0.22$ is  $0.77/0.63\approx 1.222$, while $(0.44/0.22)^{1/3} \approx 1.259$. A direct comparison with analytical predictions can be made using the  formula \cite{Por91}
\beq  \label{disprelcilD}
\frac{\pi}{2}\left(\frac{\gamma}{\bar{a} k_y}\right)^2=-\rho_s\frac{\pi}{\Delta '}+\rho_s^2 d_e \frac{\bar{a} k_y}{\gamma},
\eeq
which is valid also for large $\Delta '$.

\begin{table}[ht]
    \centering
    \caption{Comparison, for $\tau=0.1$, of  the growth rate  $\gamma$ obtained numerically and  $\gamma_T$ based on Eq. (\ref{disprelcilD}), for different values of the parameters. Relative errors are also indicated.}
    
    \begin{ruledtabular}   %pls
    \begin{tabular}{cccccc}
        \toprule
        $d_e$ & $\beta_e$ & $\gamma$ & $\gamma_T$ & Error (\%) \\
        \midrule
        0.22 & $0.02 $ & $0.63$ & $0.704$ & 10.5 \\
        0.44 & $0.005$ & $0.77$ & $0.904$ & 14.8 \\
         
        \bottomrule
    \end{tabular}
    \end{ruledtabular}   %pls
    \label{table:cilargeD}
\end{table}

The results of the comparison are summarized in Table \ref{table:cilargeD}. One can notice that the relative errors are greater than those in Table \ref{table:Prcelli}. In fact, whereas in Sec. \ref{ssec:cpsi} the values of the parameters were chosen in order to approach the asymptotic regime of validity of the theory, for the large $\Delta '$ case, the emphasis is mainly in complementing the nonlinear analysis of Sec. \ref{sec:nonlin}, for which the choice of the parameters was not dictated by adherence to analytical linear theory. Therefore, a greater discrepancy between numerical and analytical results could be expected. In particular, for the case $d_e=0.44$, we see that, for instance, the condition $\gamma \ll 1$ is not well fulfilled.

With regard to the hot-ion case, i.e. $\tau=100$, we could argue that it falls into the EMHD regime, similarly to what we did for the "constant-$\psi$" case. The most updated tearing linear theory against which we could test this, is provided, to the best of our knowledge, in Ref. \cite{Att00} and predicts, using Alfv\'en units of time, $\gamma \propto d_i d_e^{2/3}$. However, the ratio between the two growth rates of Fig. \ref{fig:largedeltaprime}, for $\tau=100$, is $6.6/3.61 \approx 1.828$, whereas, according to the theory of Ref. \cite{Att00}, their ratio should be $(20/10)\times2^{2/3} \approx 3.175$. Thus, we are evidently out of the regime of validity of the theory of Ref. \cite{Att00} and what we observe is a weaker increase of the growth rate. We remark that in Ref. \cite{Bet23} a disagreement between such theory and numerical results of direct EMHD simulations was found, suggesting that this theory is quite sensitive to the adopted regime of parameters. On the other hand, numerical results of EMHD simulations in Ref. \cite{Att00} show, in some regimes, scalings other than those predicted by the theory, and which could be more compatible with what we observe for $\tau=100$.

In summary, in  the regime of small $\Delta'$, when increasing $\tau$, we observe a compelling convergence of the numerical growth rate to the EMHD growth rate obtained from Eq.(\ref{disprelemhd}), provided condition (\ref{condemhd2}) be satisfied. Figure \ref{fig:cstpsi_2} also demonstrates a remarkable convergence, as $\beta_e$ decreases, of the numerical growth rate towards the asymptotic prediction. Differently, in the regime of large $\Delta '$ with hot ions, a noticeable deviation from the predictions of the theory of  Ref. \cite{Att00} is observed, indicating a strong sensitivity to parameter regimes, a phenomenon already mentioned in other studies \cite{Bet23}.

\section{Turbulence generated by reconnection}\label{sec:nonlin}

In this Section we discuss regimes in which Kelvin-Helmholtz instabilities, following an initial tearing instability, lead to a turbulent current layer.
The simulations considered for this study were conducted using a grid size of $2080^2$ collocation points in a 2D domain defined as $-2.2 \pi \leq x \leq 2.2\pi$ and $-1.8\pi \leq y\leq 1.8\pi$. For these runs, the characteristic length $L$ is taken equal to $ \hat{\rho}_s$. We applied filters \cite{Lel92} designed to smooth out scales for which $\hat{k}_{\perp} L = k_{\perp} > 300$. This range is well beyond the region in which $k_\perp \rho_e  \sim 1$, suggesting that the filter has a minimal impact on the spectral domain under investigation. Our primary interest concerns the dynamics at the electron inertial scale $d_e$.

The two simulations presented in this Section were performed with the following parameters:
\beq
d_e = 0.223, \quad \rho_s = 1, \quad \beta_e = 0.02, \quad d_i = 10, \quad \delta^2 = m_e/m_i = 5 \times 10^{-4}. \label{para1}
\eeq
The temperature ratio $\tau$ and $\beta_i$ are given by:
\beq
 \tau \in \{100, 1\}, \quad \beta_i \in \{2, 0.02\}. \label{para2}
\eeq

These cases correspond to an ion Larmor radius of $\rho_i= \sqrt{\tau} \rho_s =10$ and $\rho_i=1$, respectively. 

\subsection{ Case $\tau=100$}

In the case $\tau = 100$, it is worth noting that we are approaching the asymptotic regime of hot ions, as described in Section \ref{ssec:hotion}. The linear growth rate of the tearing mode is approximately three times larger than for the case $\tau = 1$,  and the maximum growth rate reached during the faster-than-exponential growth phase of the island is about five times larger. 
At the beginning of the fast growth phase, the development of the tearing mode is so rapid that the outflow directed towards the interior of the island seems to generate a mushroom-shaped structure, symptomatic of a Rayleigh-Taylor instability.  In this regime, we observe electron jets colliding and triggering turbulence. This phenomenon explains why, during the early nonlinear phase, turbulence is primarily generated at the center of the island. A similar result was previously reported in the small-$\tau$ regime \cite{Del03, Del06, Gra07, Gra09}.

\begin{figure}[h!]
    \centering
    \includegraphics[trim=0 0 0 0, scale=0.7]{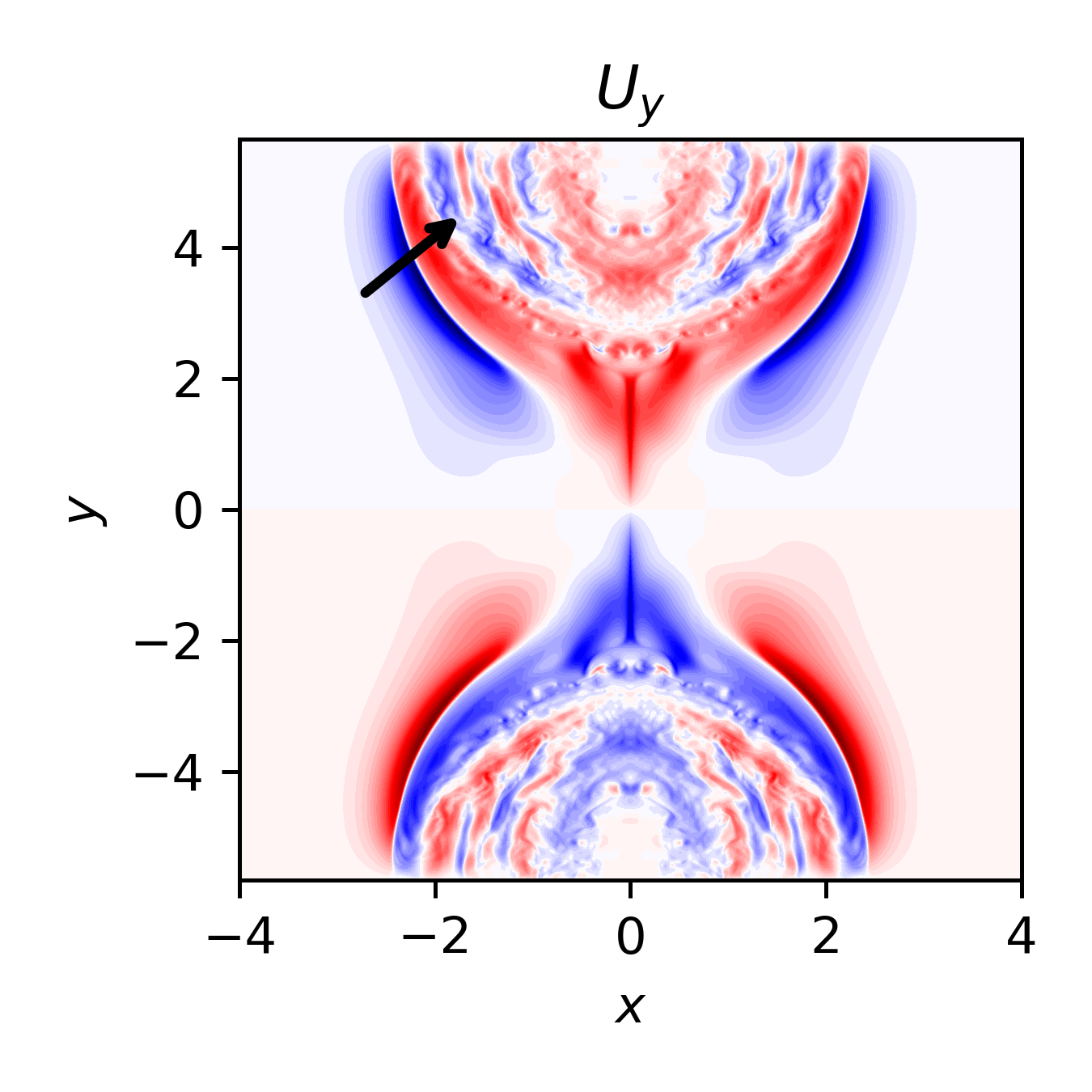}
    \includegraphics[trim=0 0 0 0, scale=0.63]{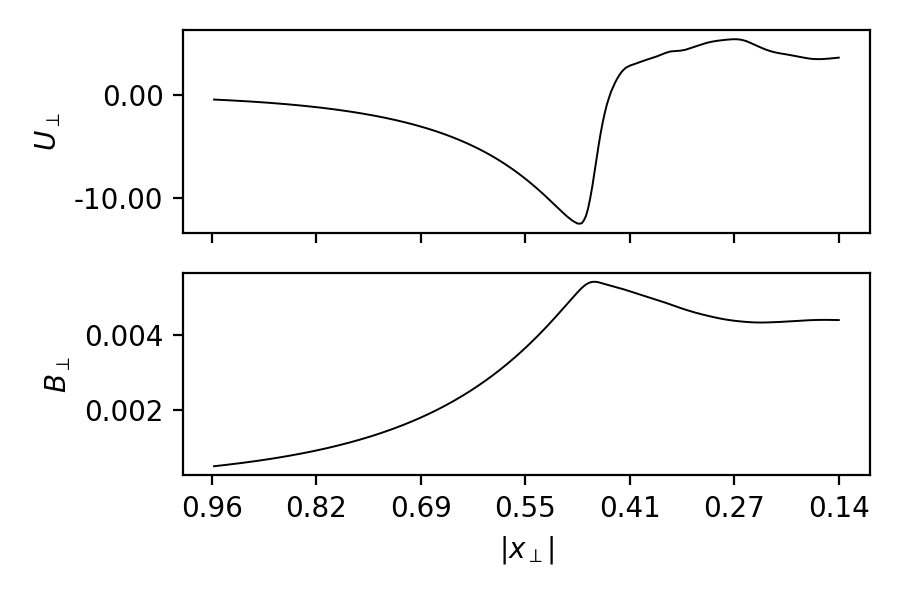}
 \caption{Left panel: Color-scale plot of $U_y$ for $\tau =100$ (red: positive values, blue: negative values). Right panel: Profiles of $U_\perp$ and $B_\perp$  along  the black arrow seen in the right panel. } 
    \label{fig:shear_125} 
\end{figure}
\begin{figure}
    \centering
\includegraphics[trim=130 0 0 0, scale=0.59]{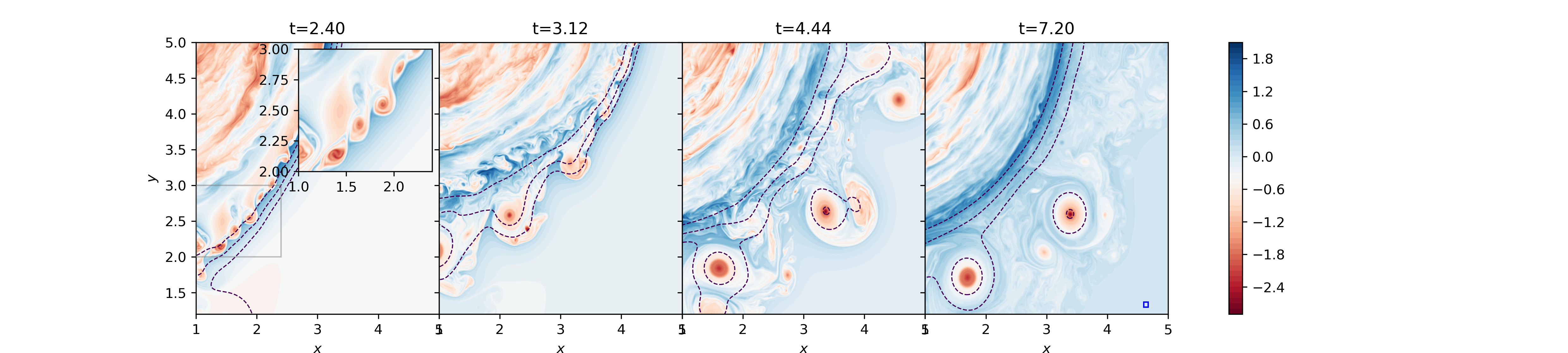}
 \caption{ Color-scale plot of $U_e$ with isolines of $A_\parallel$ showing the evolution of the turbulence outside the separatrix for $\tau=100$.  The displayed region corresponds to approximately a quarter of the total domain, focusing on the separatrix of the island. The left panel also includes a zoomed-in view within the square area delimited by  the grey line, where fluid vortices start developing. In the rightmost panel a square of size $ \pi \rho_e$ is superimposed. This shows that the size of the vortices is considerably greater than the scale at which electron FLR effects become relevant.} 
    \label{fig:ue_ev_125} 
\end{figure}

Regarding the periphery of the island, Fig. \ref{fig:shear_125}  displays a cut of the perpendicular fluid velocity $U_\perp$  and the perpendicular magnetic field $B_\perp$, across the separatrix of the island (taken along the black arrow shown on the color-scale plot). The cut reveals a localized velocity shear at the point where $B_\perp$ reaches a local maximum, corresponding to the location of the separatrix. The magnetic field lines become distorted and stretched in the direction of the shear. Subsequently, magnetic eddies form due to magnetic reconnection at the separatrices. It is worth pointing out that a strong magnetic field  which doesn't reverse across the velocity shear contributes to the stabilization of the Kelvin-Helmholtz instability.  Similar secondary fluid-like instabilities taking place at the separatrix of a magnetic island were also observed in  the context of tearing Particle-In-Cell simulations \cite{Fer12,Puc18}. In the cold-ion case, a secondary instability taking place at the separatrices was reported in Ref. \cite{Del17}. Fig. \ref{fig:ue_ev_125}  shows the out-of-plane electron gyrocenter velocity, revealing the presence of large and small-scale current structures. For comparison, we superimpose  in the right panel, a blue square with side of length $ \pi \rho_e$  equal to the half wavelength corresponding to the wavenumber $k_\perp = 1/\rho_e$.
The size of the vortices is considerably greater than the size of the square. Therefore it is reasonable to conclude that the formation of such vortices does not depend on the dynamics occurring at the scale of the electron Larmor radius, where our model is not accurate.
It is of interest to compare this run with an integration of the  2D REMHD model  \eqref{contemhd2}-\eqref{ohmemhd2}. The plots showing the time evolution of the current are presented in Fig. 6. We note a qualitative similarity with the simulation with $\tau = 100$, in particular the development of a secondary Kelvin-Helmholtz instability and the subsequent formation of magnetic vortices.
\begin{figure}
    \centering
\includegraphics[trim=130 0 0 0, scale=0.59]{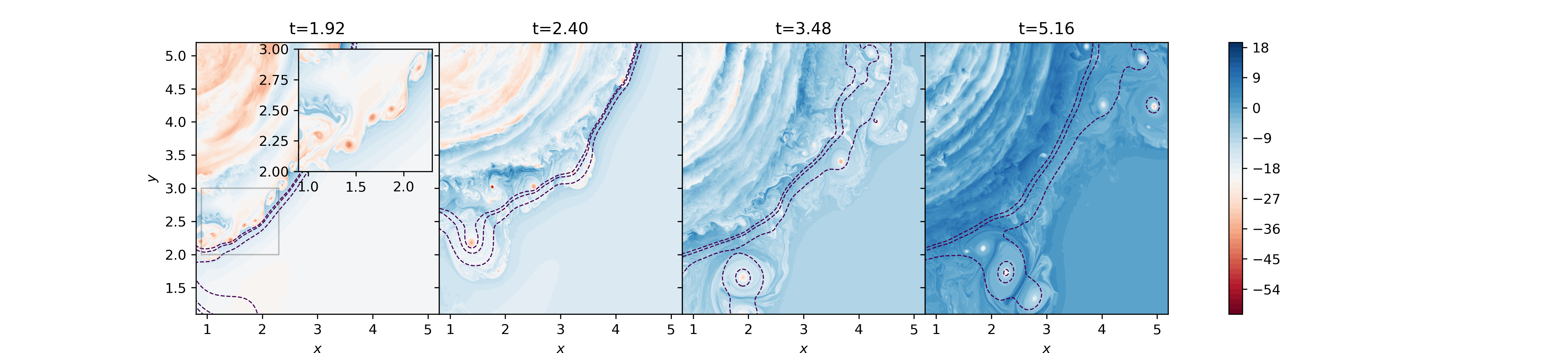}
 \caption{Same as in Fig. \ref{fig:ue_ev_125} for the REMHD simulation. }
    \label{fig:ue_ev_125EMHD} 
\end{figure}
It can be observed that the magnetic vortices (dashed black lines, corresponding to the isolines of $A_\parallel$) persist and have a similar shape in the two simulations. Figure \ref{cutes} presents cuts through a plasmoid in the parallel electron velocity for the two runs. These structures have been fitted, for comparison, with a Gaussian whose full width at half maximum is FWHM $= 0.3$ in units of $L=\hat{\rho}_s$ for both $\tau=100$ and REMHD simulations, indicating that the size of the vortices is typically $1.3 \hat{d}_e$. We see the onset of a "double structure"  especially conspicuous in the REMHD, with  the internal structure  becoming sharper when the dissipation is reduced and possibly singular in the zero-dissipation limit (not shown). This issue deserves further investigations. 

\begin{figure}
    \centering
\includegraphics[trim=10 0 0 0, scale=0.5]{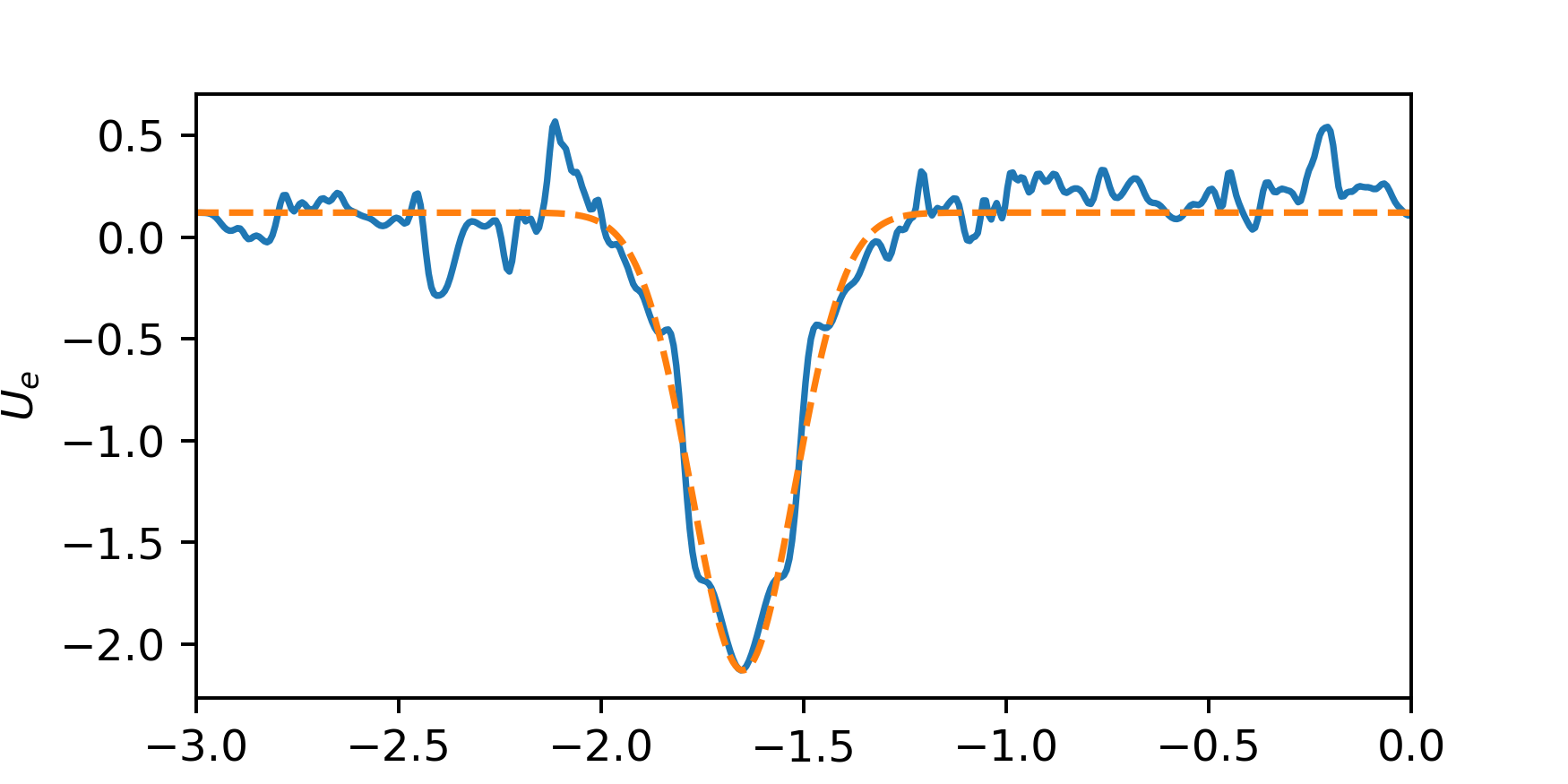}
\includegraphics[trim=10 0 0 0, scale=0.5]{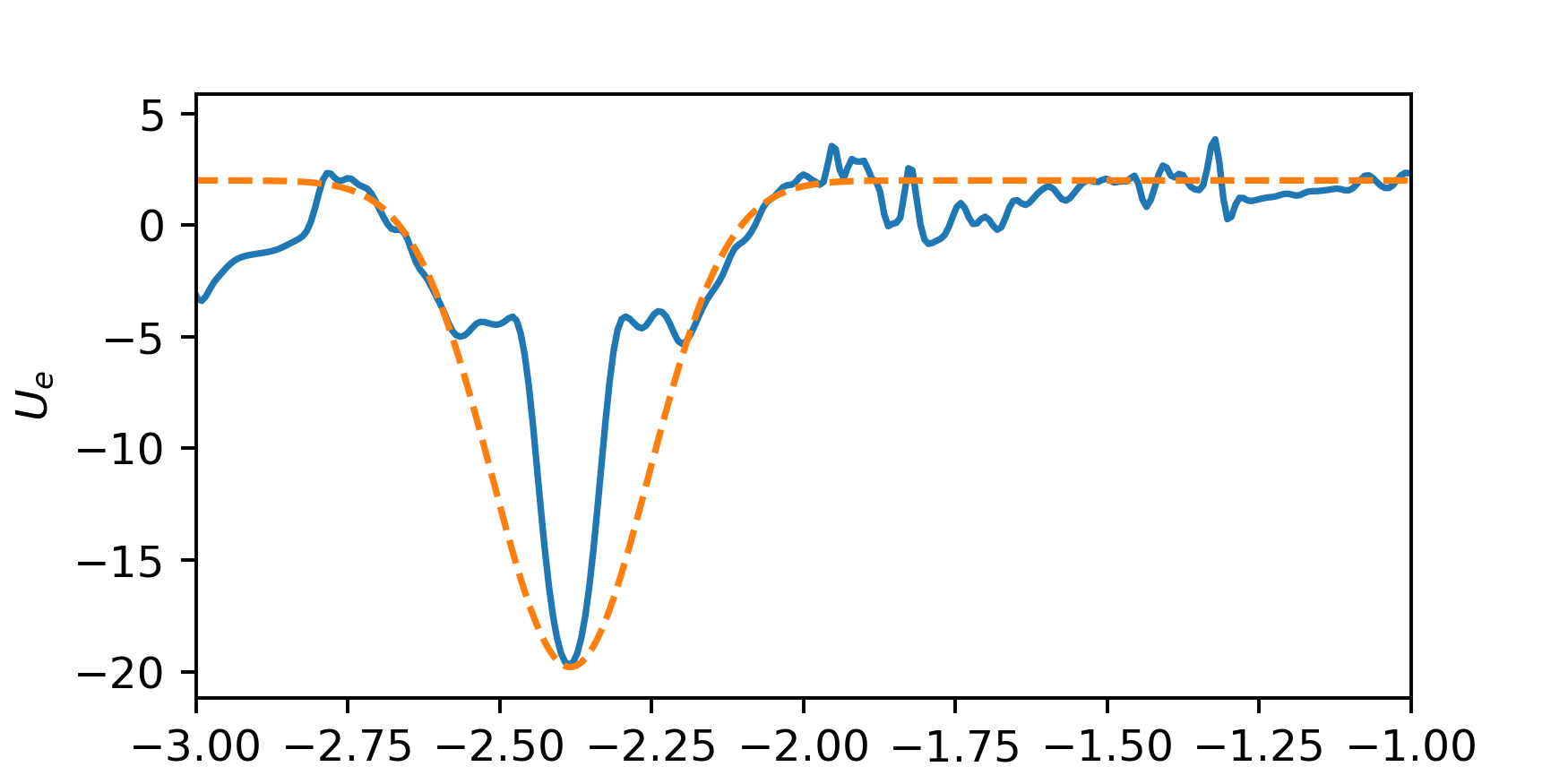}
 \caption{Cut of $U_e$ along $x$ for the run $\tau = 100$ (left panel) and for the REMHD run (right panel). Both structures are fitted by a Gaussian (orange dashed lines). }
    \label{cutes} 
\end{figure}

\begin{figure}[h!]
    \centering
   \includegraphics[trim=0 0 0 0, scale=0.6]{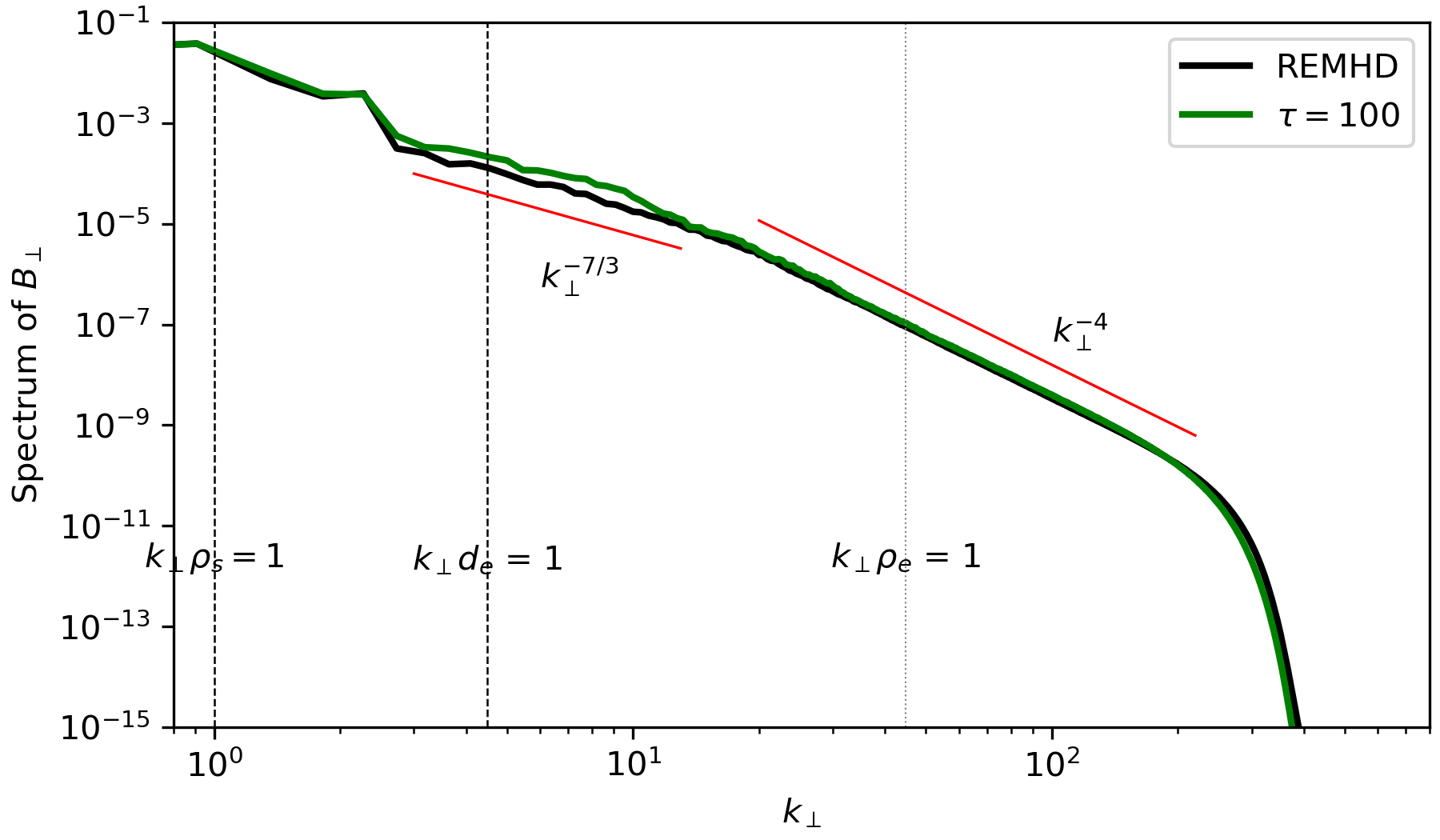}
 \caption{Left: Time averaged  spectra of the field ${\bf B}_\perp$ for $\tau=100$ and for the REMHD run. }
    \label{fig:spec125} 
\end{figure}

Time-averaged spectra of ${\bf B}_\perp$  for $\tau = 100$ and REMHD, when turbulence has become fully developed, are  plotted in Fig. \ref{fig:spec125}. On these spectra,   $k_\perp = 0.1$ corresponds to $k_\perp \rho_i = 1$ and  $k_\perp = 4.3$ to $k_\perp d_e =1$, indicating that  the simulation includes the sub-ion and sub-$d_e$ scale   ranges. At scales larger than $2d_e$, a power law consistent with a $-7/3$ exponent is observed. 
In the inertial kinetic range ($k_\perp d_e>1$), the simulation exhibits a spectral exponent close to $-4$. The slope of $-4$ is not in significant disagreement with the $-11/3 \approx -3.67$  spectral exponent predicted  for IKAW (inertial kinetic Alfvén wave) turbulence \cite{Mey10, Che17}, but the presence of structures formed by the Kelvin-Helmholtz instability can certainly impact the perpendicular magnetic spectrum slope. We note that, the spectrum in the region for $k_\perp < 10$ may be subtly influenced by the presence of the equilibrium magnetic field, while the knee appearing around $k_\perp = 300$ corresponds to the impact of filters.  Secondly,  A detailed study of this issue would preferably be performed in the framework of a homogeneous isotropic turbulence.

\subsection{ Case $\tau=1$}

\begin{figure}[h!]
    \centering
    \includegraphics[trim=0 0 0 0, scale=0.7]{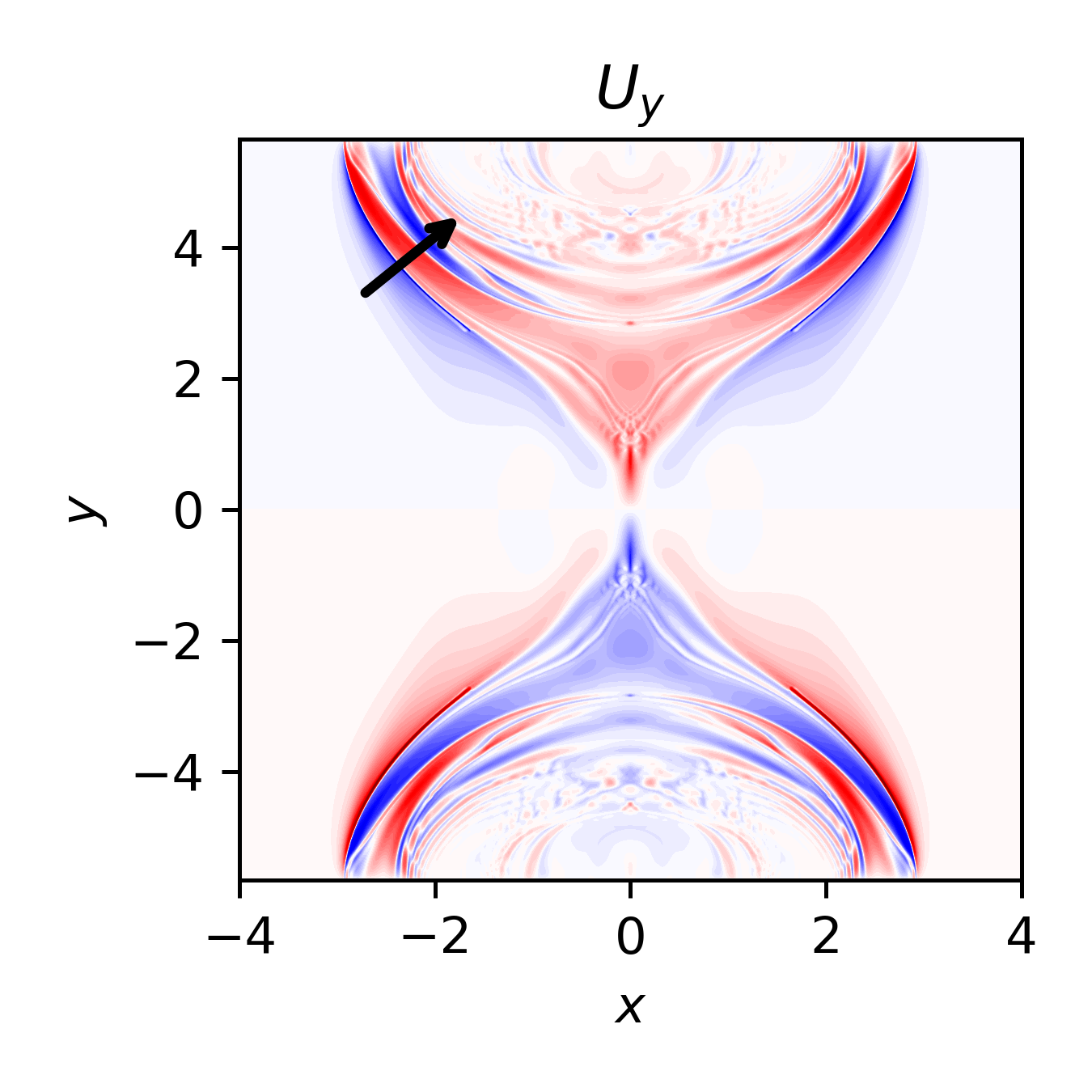}
    \includegraphics[trim=0 0 0 0, scale=0.63]{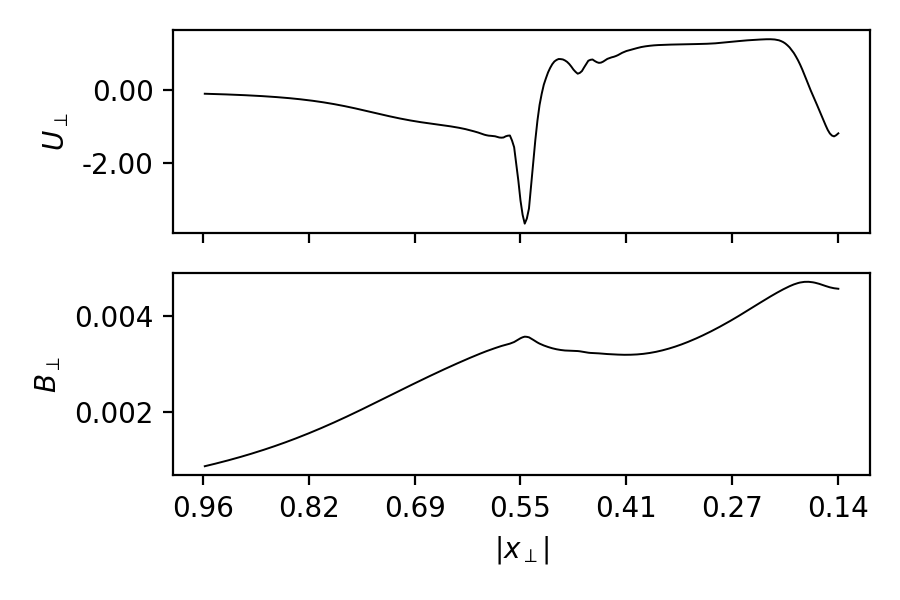}
 \caption{Left panel: Color-scale plot of $U_y$ for $\tau=1$. Right panel: Profiles of $U_\perp$ and $B_\perp$ along the black arrow visible in the right panel.}
    \label{fig:shear_122} 
\end{figure}

In the case with $\tau = 1$ (corresponding to an equilibrium current sheet with a width of the order of the ion Larmor radius),  Fig. \ref{fig:shear_122} shows that the velocity shear in $U_\perp$ is less important compared to the case $\tau = 100$ (which corresponds to a current sheet of a width $10$ times smaller than the ion Larmor radius) leading  the Kelvin-Helmholtz instability to develop at a  smaller scale. 
When considering the electron gyrocenter velocity (Fig. \ref{fig:ue_ev_125}), we observe magnetic vortices that are smaller than $d_e$. Their size indeed appears to be comparable to the wave length associated with the condition $k_\perp \rho_e = 1$ (indicated by the blue square). This raises an intriguing question: do these structures arise as a result of the electron FLR terms included in the model and which are subdominant at the scale $d_e$? To explore this problem, we conducted a simulation in which we deliberately removed the electron FLR term from the code (note that $k_\perp \rho_i$ remains unchanged in this comparison, unlike in the previous case of hot ions).  The resulting out-of-plane electron gyrocenter velocity is presented in Fig. \ref{fig:contours_tau=1_woeFLR}, highlighting a notable reduction of magnetic vortices and an absence of current structures of sizes similar to $d_e$ or smaller. On the other hand, for this run, we had to increase the range of the spectrum affected by the filter, resulting in smoother smallest scales. We can see that, for this run, the region $k_\perp > 100$ is affected by the filter.  However, given the high resolution of our simulation, there remains a sizeable range below the scale $\rho_e$. This observation could suggest the existence of a physically significant dynamics at the  scale of the electron Larmor radius, which lies beyond the scope of the present model. Retaining all electron FLR terms could be of interest for a future work, but, at finite ion temperature, it would, in this case, be essential to incorporate parallel ion dynamics into the model. For this purpose, a four-field model would be necessary.
\begin{figure}
    \centering
\includegraphics[trim=130 0 0 0, scale=0.59]{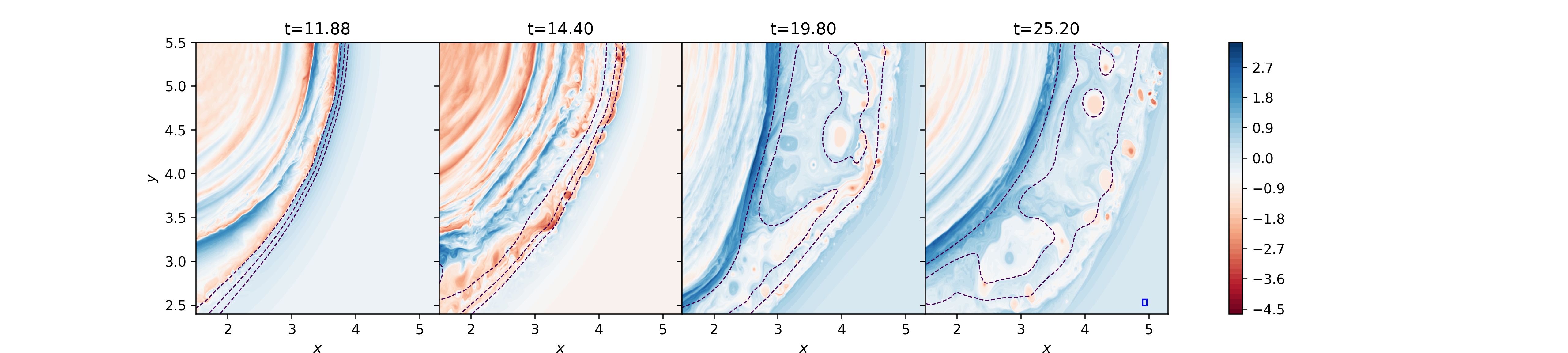}
 \caption{Same as in Fig. \ref{fig:ue_ev_125} for $\tau=1$.  }
    \label{fig:contours_tau=1} 
\end{figure}

\begin{figure}
    \centering
\includegraphics[trim=130 0 0 0, scale=0.59]{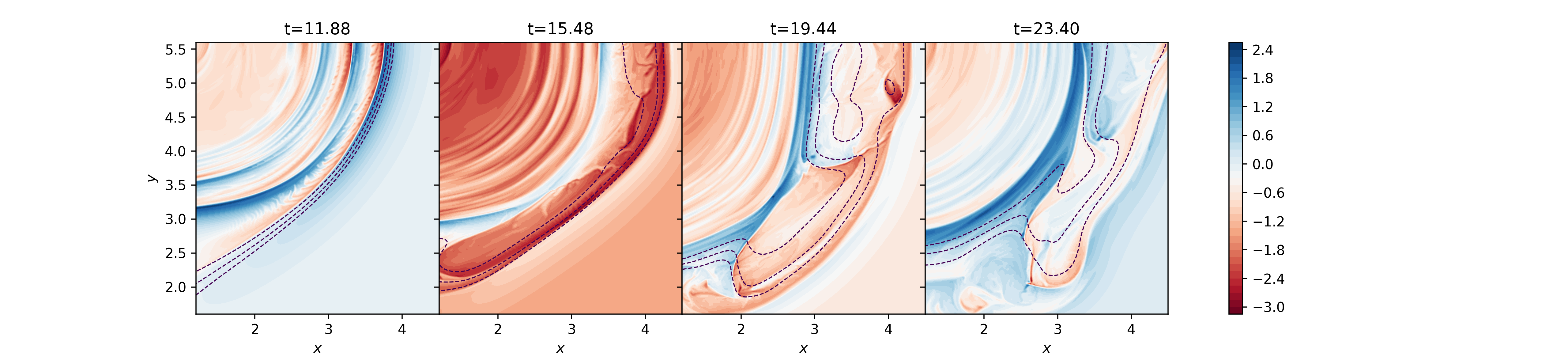}
 \caption{Same as in Fig. \ref{fig:ue_ev_125} for $\tau=1$ in the absence of the electron FLR term.  }
    \label{fig:contours_tau=1_woeFLR} 
\end{figure}
\begin{figure}[h!]
    \centering
    \includegraphics[trim=0 0 0 0, scale=0.6]{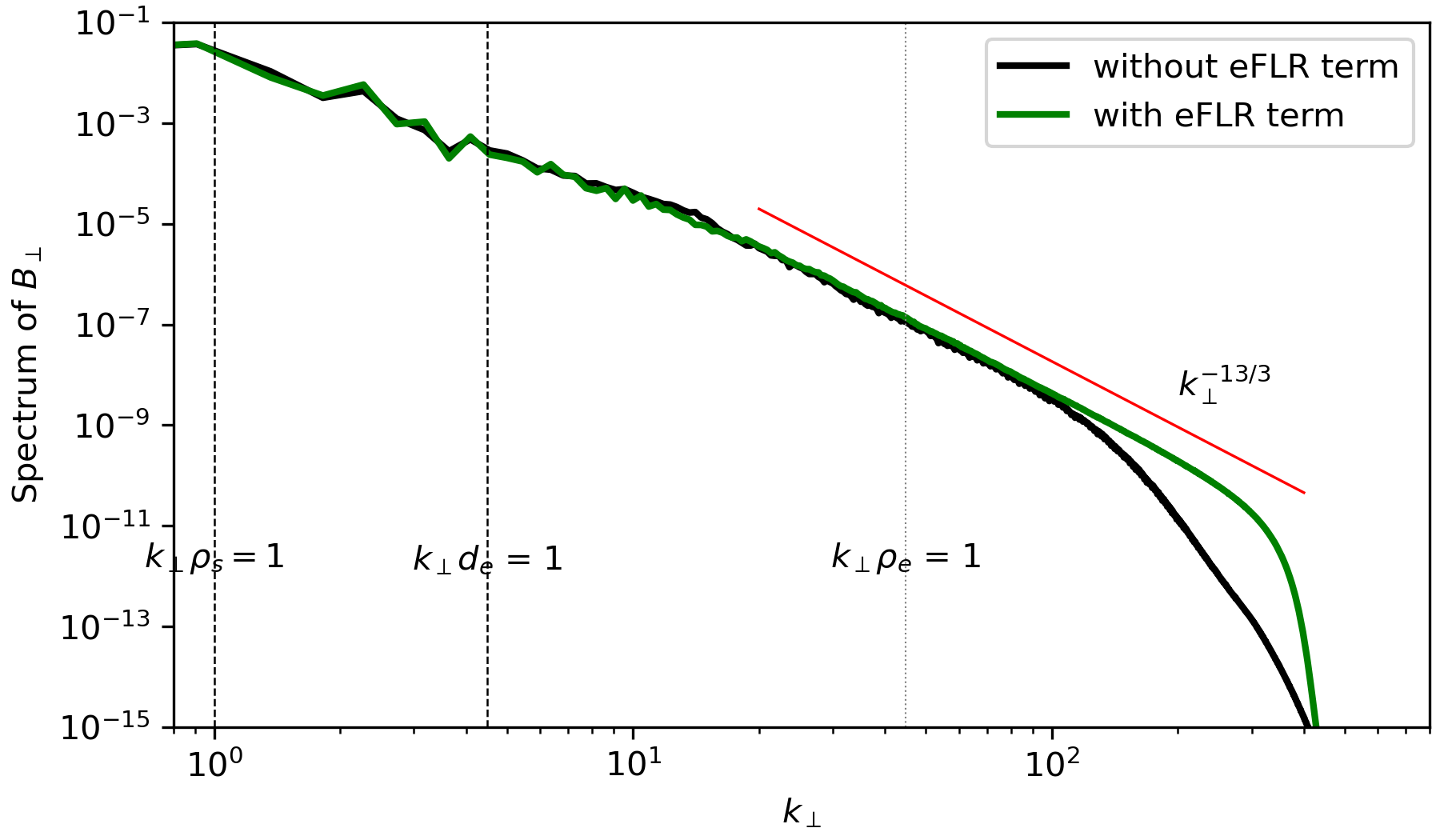}
 \caption{Time averaged  spectra of the field ${\bf B}_\perp$ for $\tau=1$.}
    \label{fig:spec122} 
\end{figure}

The time-averaged spectra of ${\bf B}_\perp$  for the simulations with and without electron FLR corrections, after turbulence has fully developed are shown in Fig. \ref{fig:spec122}. On these spectra, $k_{\perp \rho_i}=2\pi/\rho_i$ is 0.62 and $k_{\perp_{d_e}}=2\pi/d_e$ is 4.3, indicating that  the simulations cover the sub-ion and sub-$d_e$ scale spectral ranges.  A power law $E_{B_\perp}(k_\perp) \propto k_\perp^{-13/3}$ is observed below $k_{\perp_{d_e}}$.

\section{Conclusions}\label{sec:conclusion}

In this study, we have investigated the linear and nonlinear evolution of the tearing mode instability in a collisionless plasma for different ion-to-electron temperature ratios. Through numerical simulations, we  explored the behavior of the instability in different regimes, focusing on cases with moderate and large ion Larmor radii compared to the characteristic width of the equilibrium current sheet.
In particular we studied the transition of the instability behavior from a cold-ion to an EMHD regime as $\tau$ increases. 
In the small $\Delta'$ regime, we numerically validated analytical dispersion relations  and showed how the growth rate evolves from a linear scaling in $d_e \Delta'$ for cold ions  \cite{Porcelli02} to a  scaling in $d_e^2 \Delta^{'2}$ for hot ions \cite{Bul92}. As expected, the agreement with the analytical EMHD growth rate requires the width of the inner region be greater than the electron thermal Larmor radius. 

In  the regimes  $\tau = 100$ and $\tau = 1$, we examined the development of turbulence triggered by the Kelvin-Helmholtz instabilities in the nonlinear phase of the tearing mode, concentrating on the influence of the ion-to-electron temperature ratio on the development of turbulence. 

For $\tau = 100$, the presence of large magnetic vortices, formed as a consequence of the Kelvin-Helmholtz instability, is noted. The size of such vortices is much larger than the scale at which electron FLR effects (which are not consistently taken into account in our model) become relevant. We can then assume that their formation is genuine and not influenced by deficiencies of the model. We observed analogous vortices also in EMHD simulations, where electron FLR effects are absent. At a closer inspection, EMHD vortices exhibit a "double structure", with an inner core whose width appears to be influenced by dissipative effects.  The turbulent spectrum exhibits power law of $-4$, which requires further investigation, preferably in the framework of homogeneous and isotropic turbulence. 

For $\tau = 1$, and $\beta_e = 0.028$ we also observe a secondary Kelvin-Helmholtz-like instability with subsequent formation of vortices. However, in this case, the electron FLR term present in our model appears to play a more important role in their formation. Concerning the spectrum, a power law close to $-13/3$ is observed. In this regime, the presence of small-scale structures, their reduction upon the removal of the electron FLR term, and the unexplored dynamics at scales below $\rho_e$ emphasize that further explorations, with the inclusion of all the electron FLR effects and the parallel ion dynamics, could unlock valuable insights. 

Future developments of this study also include  simulations in three space dimensions, where a richer and more complex dynamics is expected. Also in the perspective of including the dynamics of higher order moments and kinetic effects, the 2D assumption can be rather restrictive. For instance, as shown in Ref. \cite{Tas18}, kinetic effects on the tearing growth rate, described by means of a Landau closure,  become appreciable only in 3D. Furthermore, investigating  reconnection arising during the evolution  of a freely-decaying turbulence would allow us to explore the self-organization of the plasma in the absence of external driving. These extensions will contribute to a deeper understanding of the interplay between reconnection and turbulence in magnetized plasmas.

\section*{Acknowledgments}

The Authors are very grateful to Daniele Del Sarto and to Homam Betar for fruitful discussions and for having carried out for us numerical tests for EMHD tearing growth rates. The computations have been done on the “Mesocentre SIGAMM” machine, hosted by Observatoire de la C\^ote d’Azur.

%\newpage
%\section{Additional}

\bibliographystyle{plainnat}

%\selectbiblanguage{french}

%\bibliography{flr}
\bibliography{recsubion}

\end{document}